\documentclass[
aps,nofootinbib,showpacs,showkeys,preprint
,tightenlines ] {revtex4}

\usepackage{epsf,epsfig,subfigure,graphicx,amsmath,amssymb}
\usepackage{color}

\newcommand{\dis}[1]{\begin{equation}\begin{split}#1\end{split}\end{equation}}
\newcommand{\be}{\begin{equation}}
\newcommand{\ee}{\end{equation}}
\def\bea{\begin{eqnarray}}
\def\eea{\end{eqnarray}}

\newcommand{\eq}[1]{Eq.~(\ref{#1})}

\newcommand{\VEV}[1]{\langle #1 \rangle}

\newcommand{\tev}{\,\textrm{TeV}}
\newcommand{\gev}{\,\textrm{GeV}}

\newcommand{\ev}{\,\textrm{eV}}

\def\tb{\tan\beta}

\begin{document}


\title{\Large\bf Precise focus point scenario 
\\for a natural Higgs boson in the MSSM 
}

\author{Bumseok Kyae$^{(a)}$\footnote{email: bkyae@pusan.ac.kr}
and Chang Sub Shin$^{(b)}$\footnote{email: changsub@physics.rutgers.edu}
}
\affiliation{
$^{(a)}$ Department of Physics, Pusan National University, 
Busan 609-735, Korea 
\\
$^{(b)}$ Department of Physics and Astronomy, New High Energy Theory Center, 
Rutgers University, Piscataway, New Jersey 08854, USA
}

\begin{abstract}

A small Higgs mass parameter $m_{h_u}^2$ 
can be insensitive to various trial heavy stop masses, 
if a universal soft squared mass is assumed for the chiral superpartners and the Higgs boson at the grand unification (GUT) scale, and a focus point (FP) of $m_{h_u}^2$  
appears around the stop mass scale. 
The challenges in the FP scenario are (1) a too heavy stop mass ($\approx 5 \tev$) needed for the 126 GeV Higgs mass and (2) the too high gluino mass bound ($\gtrsim 1.4 \tev$).  
For a successful FP scenario, 
we consider (1) a superheavy right-hand (RH) neutrino
and (2) the first and second generations of hierarchically heavier chiral superpartners. 
The RH neutrino can move a FP in the higher energy direction in the space of $(Q, ~m_{h_u}^2(Q))$, where $Q$ denotes the renormalization scale.   
On the other hand, the hierarchically heavier chiral superpartners can lift up a FP in that space through two-loop gauge interactions.
Precise focusing of $m_{h_u}^2(Q)$ is achieved with the RH neutrino mass of $\sim 10^{14} \gev$ 
together with an order one ($0.9-1.2$) Dirac Yukawa coupling to the Higgs boson, 
and the hierarchically heavy masses of $15-20 \tev$ for the heavier generations of superpartners, when the U(1)$_R$ breaking soft parameters, $m_{1/2}$ and $A_0$ are set to be $1 \tev$ at the GUT scale. 
Those values can naturally explain the small neutrino mass through the seesaw mechanism, and suppress the flavor violating processes in supersymmetric models.

%
%

\end{abstract}




\maketitle


\section{Introduction} 


The naturalness problem of the electroweak scale (EW) and the Higgs boson mass has been 
the most important issue for the last four decades in the theoretical particle physics community.  
It has provided a strong motivation 
to study various theories beyond the standard model (SM). 
Particularly, the minimal supersymmetric SM (MSSM) has been regarded as the most promising candidate 
among new physics models beyond the SM. 
However, any evidence of new physics beyond the SM including supersymmetry (SUSY) has not been observed yet at the large hadron collider (LHC), 
and experimental bounds on SUSY particles are increasing gradually. 
Nonetheless, a better new idea that can replace the present status of SUSY has not seemed  to appear yet.  
Accordingly, it would be worthwhile to explore a breakthrough within the SUSY framework.

Concerning the radiative Higgs mass and EW symmetry breaking, the top quark Yukawa coupling ($y_t$) of order unity plays the key role in the MSSM:
through the sizable top quark Yukawa coupling, 
the top quark and stop make a dominant contribution to the renormalization of the soft mass parameter of the Higgs boson ($m_{h_u}^2$) as well as the radiative physical Higgs mass squared ($m_H^2$) \cite{book}:  
\bea \label{physHiggs}
&&\Delta m_H^2\approx 
\frac{3y_t^4}{4\pi^2}{\rm sin}^4\beta v_h^2 ~{\rm log}\left(\frac{\widetilde{m}_t^2}{m_t^2}\right) 
+ \cdots , 
\\  \label{renormHiggs}
&&  
~~
\Delta m_{h_u}^2\approx \frac{3y_t^2}{8\pi^2}\widetilde{m}_t^2 
~{\rm log}\left(
\frac{\widetilde{m}_t^2}{\Lambda^2}\right) +\cdots ,  
\eea
where $m_t$ ($\widetilde{m}_t$) denotes the top quark (stop) mass, and $v_h$ 
is the vacuum expectation value (VEV) of the Higgs boson, $v_h\equiv\sqrt{\VEV{h_u}^2+\VEV{h_d}^2}\approx 174~{\rm GeV}$  with $\tan\beta\equiv\VEV{h_u}/\VEV{h_d}$. 
%
$\Lambda$ means a cutoff scale. A messenger scale of SUSY breaking is usually adopted for it. 
Here we set the left-hand (LH) and right-hand (RH) stop squared masses, $m_{q_3}^2$ and $m_{u^c_3}^2$ equal to $\widetilde{m}_t^2$ for simplicity.  
Note that $\Delta m_{h_u}^2$ can be a large negative value for 
a large stop mass and a high messenger scale. 

As seen in \eq{physHiggs}, a large stop mass can raise the radiative Higgs mass. 
According to the recent analysis based on three-loop calculations \cite{3-loop}, 
a $3$--$4$ or 
$5 \tev$ stop mass is necessary for explaining the recently observed 126 GeV Higgs mass \cite{LHCHiggs} without a stop mixing effect. 
From \eq{renormHiggs}, however, such a heavy stop mass is expected to significantly enhance the renormalization effect on $m_{h_u}^2$, and 
eventually it gives rise to a fine-tuning problem associated with naturalness of the EW scale. 
It is because a negative $m_{h_u}^2$ triggers the EW symmetry breaking, and eventually determines the $Z$ boson mass in the MSSM, as seen in the extremum condition of the MSSM Higgs potential \cite{book}:  
\dis{ \label{extremeCondi}
\frac12 m_Z^2=\frac{m_{h_d}^2-m_{h_u}^2{\rm tan}^2\beta}{{\rm tan}^2\beta-1}
-|\mu|^2 , 
}
where $m_Z^2$ denotes the $Z$ boson mass and $\mu$ is the ``$\mu$-term'' coefficient in the MSSM superpotential. 
If $-m_{h_u}^2$ is excessively large, it should be compensated with $|\mu|^2$. 
Thus, a fine-tuning of $10^{-3}$--$10^{-4}$ does not seem to be avoidable in the MSSM, 
unless the messenger scale $\Lambda$ is low enough. 
Due to this reason, a relatively smaller stop mass ($\ll 1 \tev$) has been assumed 
for naturalness of the EW scale, and various extensions of the Higgs sector have been proposed for explaining the observed 126 GeV Higgs mass \cite{NMSSMreview,singletEXT,extensions}. 
Unfortunately, however, the stop mass bound has already reached $700 \gev$ \cite{stopmass}, 
which starts threatening the traditional status of SUSY as a solution to the naturalness  problem of the EW phase transition. 
Thus, in this paper, we intend to discuss the naturalness problem in case the stop is quite heavy ($\sim 5 \tev$).

In fact, the renormalization of $m_{h_u}^2$, \eq{renormHiggs} is necessarily affected by ultraviolet (UV) physics. 
Thus, for a more complete expression of it, the full renormalization group (RG) equations 
should be studied for a given UV model, 
even though \eq{renormHiggs} would not be very sensitive to an UV physics in SUSY models.  
Unlike the expectation based on low energy physics, however, 
it was claimed that 
the $Z$ boson and Higgs masses at low energy are quite insensitive to the stop mass in the ``focus point (FP) scenario'' \cite{FMM1,FMM2,Nath}: 
%
under the simple initial condition for the stops and Higgs squared masses, 
$m_{q_3}^2=m_{u^c_3}^2=m_{h_u}^2=\cdots\equiv m_0^2$ 
at the grand unification (GUT) scale, 
the RG solution of $m_{h_u}^2$ turns out to be almost independent of $m_0^2$ 
{\it at the EW scale} unlike those of $m_{q_3}^2$ and $m_{u^c_3}^2$. 
It is because the coefficient of $m_0^2$ in the RG solution of $m_{h_u}^2$ at the EW scale
turns out to be quite small. 
Accordingly, $m_{h_u}^2$ can remain small enough even for relatively large trial $m_0^2$s ($\sim$ multi-TeV)  
unlike other superparticles in the chiral sector.  
Interestingly enough, moreover, the FP scenario favors the simplest version of SUSY model with the minimal field contents and the universal initial condition for the soft squared masses at the GUT scale:     
many careless extensions of the MSSM at low energy would destroy the FP mechanism.

The insensitivity of $m_{h_u}^2$ to $m_0^2$ or stop masses implies 
that \eq{renormHiggs} is effectively canceled by other ingredients.
One might expect that a fine-tuning for smallness of $m_{h_u}^2$ 
would be hidden somewhere in this scenario. 
This guess is actually true. 
As will be seen later, the smallness of the coefficient of $m_0^2$ in $m_{h_u}^2$ originates from the {\it fact} that 
\dis{ \label{key}
e^{\frac{-3}{4\pi^2}\int^{t_0}_{t_W}dt ~y_t^2}\approx \frac13 ~.
}
Here $t$ parametrizes the renormalization scale $Q$, $t-t_0={\rm log}\frac{Q}{M_G}$. 
$t_W$ and $t_0$ correspond to the EW and GUT scale $M_G$ ($\approx 2\times 10^{16} \gev$), respectively.      
Actually, \eq{key} is an accidental relation in some sense. 
Just the quark and lepton masses, the low energy values of the SM gauge couplings, 
and the MSSM field contents completely determine $y_t(t)$, and 
the $Z$ boson mass scale and the gauge coupling unification scales 
provide exactly the needed energy interval. 
In the sense that \eq{key} is not artificially designed, but Nature might permit it,   
we will call it ``Natural tuning.'' 
Of course, there might exist a deep reason for it. 
In this paper, however, we will not attempt to explain the origin, 
but take a rather pragmatic attitude: we will just accept, utilize, and improve it.

However, the recently observed 126 GeV Higgs mass is challenging also in the FP scenario. 
Since the FP scenario works well with the minimal field contents and a suppressed stop mixing effect, 
the Higgs mass can be raised only through the radiative correction by the quite heavy stop, 
$\widetilde{m}_t\sim 3$--$4$ or $5 \tev$ \cite{3-loop}. 
To get a heavier stop mass, we need a larger $m_0^2$. 
In order for $m_{h_u}^2$ to remain insensitive 
even to much larger $m_0^2$s [$>(5 \tev)^2$], 
a more precise focusing is quite essential. 
That is to say, the coefficient of $m_0^2$ in the $m_{h_u}^2$'s RG solution 
should be much closer to zero. 
Moreover, $m_{h_u}^2$ does not follow the original FP scenario below the stop mass scale, 
because the stops are decoupled there. 
Thus, for a predictive EW scale, the FP should appear around the stop mass scale rather than the conventional EW or $Z$ boson mass scale.   
The present heavy gluino mass bound at the LHC, $M_3\gtrsim 1.4 \tev$ \cite{gluinomass}, 
also spoils the success of the FP scenario \cite{M3/M2,M3/M2model,DQW}.
The heavy gluino leads to  
a too large negative $m_{h_u}^2$ at the EW scale through RG evolution. 
Such an RG effect by a heavy gluino mass should be compensated properly 
for a small enough $Z$ boson mass. 

In this paper, we will attempt just to trim the FP scenario such that the FP is made located around the stop mass scale and the heavy gluino effect becomes mild. 
In order to accomplish that goal, we will consider a superheavy RH neutrino  \cite{RHnuFP0,RHnuFP}, and the two-loop {\it gauge} interactions by the hierarchically heavier first and second generations of 
chiral superpartners (sfermions) \cite{Natural/2-loop,Natural/splitZprime}.  
Hierarchically heavy masses for the first two generations of sfermions
($\gtrsim 15 \tev$) could also sufficiently suppress unwanted SUSY flavor 
and SUSY $CP$ violating processes 
as in the ``effective SUSY model'' \cite{effSUSY}.    
Once the location of the FP is successfully modified to a desirable position, 
even a quite heavy stop mass could still be naturally compatible with the $Z$ boson mass scale, and the 126 GeV Higgs mass can be supported dominantly by the radiative correction from such a heavy stop. 

This paper is organized as follows: 
we will review the FP scenario and discuss the problems associated with the recent experimental results in Sec. \ref{sec:FP}. 
In Sec. \ref{sec:preciseFP}, we will explore the ways to move the location of the FP 
into a desirable position 
in the space of $(Q, ~m_{h_u}^2(Q))$. 
In Sec. \ref{sec:model}, we will propose a simple model and discuss phenomenological constraints. 
Section \ref{sec:conclusion} will be a conclusion. 
For convenience, in our discussion in the main text, 
we will leave the details of the full RG equations and derivation of some semianalytic solutions to them in the Appendix.



\section{Focus point scenario}  \label{sec:FP}


Based on our semianalytic solutions to the RG equations, 
let us discuss first the RG behaviors of soft parameters associated with the Higgs boson and the third generation of sfermions. 
When $\tb$ is small enough, the top quark Yukawa coupling, $y_t$ dominantly drives the RG running of $\{m_{h_u}^2, m_{u^c_3}^2, m_{q_3}^2, A_t\}$, 
while the bottom quark and tau lepton's Yukawa couplings, $y_b$ and $y_\tau$ are safely ignored.  
Here, $A_t$ denotes the ``$A$-term'' coefficient corresponding to the top quark Yukawa coupling.
Thus, for small $\tb$, the one-loop RG equations for $\{m_{h_u}^2, m_{u^c_3}^2, m_{q_3}^2, A_t\}$ are written as   
\bea
16\pi^2\frac{d}{dt}m_{h_u}^2&=&6y_t^2\left(X_t+A_t^2\right)-6g_2^2M_2^2-\frac65 g_1^2M_1^2 , 
\label{RG1} \\
16\pi^2\frac{d}{dt}m_{u^c_3}^2&=&4y_t^2\left(X_t+A_t^2\right)-\frac{32}{3}g_3^2M_3^2-\frac{32}{15} g_1^2M_1^2 ,
\label{RG2} \\
16\pi^2\frac{d}{dt}m_{q_3}^2&=&2y_t^2\left(X_t+A_t^2\right)-\frac{32}{3}g_3^2M_3^2-6g_2^2M_2^2-\frac{2}{15} g_1^2M_1^2 ,
\label{RG3} \\
8\pi^2\frac{d}{dt}A_t&=&6y_t^2A_t-\frac{16}{3}g_3^2M_3-3g_2^2M_2-\frac{13}{15} g_1^2M_1 ,
\label{RG4}
\eea
where $X_t\equiv m_{h_u}^2+m_{u^c_3}^2+m_{q_3}^2$. 
$t$ parametrizes the renormalization scale $Q$, $t-t_0={\rm log}\frac{Q}{M_{G}}$. 
$g_{3,2,1}$ and $M_{3,2,1}$ in the above equations stand for the three MSSM gauge couplings and gaugino masses.  
Our semianalytic solutions to them are approximately given by 
\bea
&&m_{h_u}^2(t)\approx m_{h_u0}^2+\frac{X_0}{2}\left[e^{\frac{3}{4\pi^2}\int^t_{t_0}dt^\prime y_t^2}-1\right]
+\frac{F(t)}{2}
-\frac32\left(\frac{m_{1/2}}{g_0^2}\right)^2\left\{g_2^4(t)-g_0^4\right\} ,
\label{RGsol1} \\
&&m_{u^c_3}^2(t)\approx m_{u^c_30}^2+\frac{X_0}{3}\left[e^{\frac{3}{4\pi^2}\int^t_{t_0}dt^\prime y_t^2}-1\right]
+\frac{F(t)}{3}
+\frac89\left(\frac{m_{1/2}}{g_0^2}\right)^2\left\{g_3^4(t)-g_0^4\right\} ,
\label{RGsol2} \\
&&m_{q_3}^2(t)\approx m_{q_30}^2+\frac{X_0}{6}\left[e^{\frac{3}{4\pi^2}\int^t_{t_0}dt^\prime y_t^2}-1\right]
+\frac{F(t)}{6}
+\left(\frac{m_{1/2}}{g_0^2}\right)^2\left\{\frac89 g_3^4(t)
-\frac32 g_2^4(t)+\frac{11}{18}g_0^4\right\} ,
~~~~~ \label{RGsol3} \\ 
&&A_t(t)=e^{\frac{3}{4\pi^2}\int^t_{t_0}dt^\prime y_t^2} 
\left[A_0-\frac{1}{8\pi^2}\int^t_{t_0}dt^\prime G_A
e^{\frac{-3}{4\pi^2}\int^{t'}_{t_0}dt^{\prime\prime} y_t^2}
\right] , 
\label{RGsol4}
\eea
where we ignored the bino mass $M_1$ and the relevant U(1)$_Y$ gauge contributions due to their smallness.  
For the complete expressions and derivation of the above solutions, refer to the Appendix  (setting $\widetilde{m}^2=0$).
Here, $\{m_{h_u0}^2, m_{u^c_30}^2, m_{q_30}^2, A_0\}$ denote the values of 
$\{m_{h_u}^2(t), m_{u^c_3}^2(t), m_{q_3}^2(t), A_t(t)\}$ at the GUT scale, 
and $X_0\equiv  m_{h_u0}^2+m_{u^c_30}^2+m_{q_30}^2$.   
$g_0$ and $m_{1/2}$ are the unified gauge coupling and gaugino mass at the GUT scale, 
respectively. 
$F(t)$ in the above solutions is defined as 
\dis{ \label{F}
&\quad~~ F(t)\equiv \frac{3}{4\pi^2} ~e^{\frac{3}{4\pi^2}\int^t_{t_0}dt^\prime y_t^2} \int^t_{t_0}dt^\prime ~y_t^2A_t^2 ~e^{\frac{-3}{4\pi^2}\int^{t'}_{t_0}dt^{\prime\prime}y_t^2}
\\
&-\frac{1}{4\pi^2} \left[e^{\frac{3}{4\pi^2}\int^t_{t_0}dt^\prime y_t^2} \int^t_{t_0}dt^\prime ~G_X^2 ~e^{\frac{-3}{4\pi^2}\int^{t'}_{t_0}dt^{\prime\prime}y_t^2}
-\int^t_{t_0}dt^\prime~G_X^2 \right] .
}
$G_A$ in \eq{RGsol4} and $G_X^2$ in \eq{F} are given, respectively, by  
\begin{eqnarray} \label{GA}
&&G_A(t)\equiv \frac{16}{3}g_3^2M_3+3g_2^2M_2+\frac{13}{15}g_1^2M_1 
=\left(\frac{m_{1/2}}{g_0^2}\right)\left[\frac{16}{3}g_3^4+3g_2^4+\frac{13}{15}g_1^4\right] ,
\\ \label{GX2}
&&G_X^2(t)\equiv \frac{16}{3}g_3^2M_3^2+3g_2^2M_2^2+\frac{13}{15}g_1^2M_1^2 
= \left(\frac{m_{1/2}}{g_0^2}\right)^2\left[\frac{16}{3}g_3^6+3g_2^6+\frac{13}{15}g_1^6\right] .  
\end{eqnarray}
Note that $F(t)$ is independent of $\{m_{h_u0}^2, m_{u^c_30}^2, m_{q_30}^2\}$, so $\{m_{h_u0}^2, m_{u^c_30}^2, m_{q_30}^2\}$ appear 
only in the first three terms in the above RG solutions, Eqs.~(\ref{RGsol1}), (\ref{RGsol2}), and (\ref{RGsol3}). 

$F(t)$ depends on $\tb$ in principle. 
But it turns out to be almost insensitive to $\tb$. 
For instance, $F(t)$ at $Q=5 \tev$ [$=F(t_T)$] is estimated as 
\dis{ \label{F(t_T)}
F(t_T)\approx \left\{-1.03,-1.02\right\}\times \left(\frac{m_{1/2}}{g_0^2}\right)^2 
} 
for $\{\tb=5,\tb=50\}$ and $A_0=0$. 
Here the numerical estimation for $\tb=50$ was performed by including $y_b$ and $y_\tau$ effects with $m_{h_d}^2=m_{e^c_3}^2=m_{l_3}^2=m_0^2$.
For the complete RG equation we used, see the Appendix. 
Thus, the last three terms in \eq{RGsol1} at $Q=5 \tev$ yield $\{-1.43,-1.41\}\times m_{1/2}^2$ for $\{\tb=5,\tb=50\}$ and $A_0=0$. 
Note that the $F(t)$ term dominates over the last two terms in \eq{RGsol1} at $Q=5 \tev$. 
Although the last two terms provide a positive coefficient of $m_{1/2}^2$,   
the large gluino mass effect contained in $F(t)$ flips the sign.
%

If the gauge sector's contributions proportional to $m_{1/2}^2$ are relatively suppressed, $A_t(t)$ and $F(t)$ are simplified as follows: 
\bea
A_t(t)\approx A_0e^{\frac{3}{4\pi^2}\int^t_{t_0}dt^\prime y_t^2} , 
~~~{\rm and}~~~F(t)\approx A_0^2 ~e^{\frac{3}{4\pi^2}\int^t_{t_0}dt^\prime y_t^2}  
\left[e^{\frac{3}{4\pi^2}\int^t_{t_0}dt^\prime y_t^2}-1\right] .
\eea 
In this case, $\{m_{h_u}^2(t), m_{u^c_3}^2(t), m_{q_3}^2(t)\}$ thus reduce to 
\bea \label{appxH}
&&m_{h_u}^2(t)\approx m_{h_u0}^2+\frac{X_0}{2}\left[e^{\frac{3}{4\pi^2}\int^t_{t_0}dt^\prime y_t^2}-1\right]
+\frac{A_0^2}{2} e^{\frac{3}{4\pi^2}\int^t_{t_0}dt^\prime y_t^2} \left[e^{\frac{3}{4\pi^2}\int^t_{t_0}dt^\prime y_t^2}-1\right] + \cdots ,
\qquad \\ \label{appxU}
&&m_{u^c_3}^2(t)\approx m_{u^c_30}^2+\frac{X_0}{3}\left[e^{\frac{3}{4\pi^2}\int^t_{t_0}dt^\prime y_t^2}-1\right]
+\frac{A_0^2}{3} e^{\frac{3}{4\pi^2}\int^t_{t_0}dt^\prime y_t^2} \left[e^{\frac{3}{4\pi^2}\int^t_{t_0}dt^\prime y_t^2}-1\right]  + \cdots ,
\\ \label{appxQ}
&&m_{q_3}^2(t)\approx m_{q_30}^2+\frac{X_0}{6}\left[e^{\frac{3}{4\pi^2}\int^t_{t_0}dt^\prime y_t^2}-1\right]
+\frac{A_0^2}{6} e^{\frac{3}{4\pi^2}\int^t_{t_0}dt^\prime y_t^2} \left[e^{\frac{3}{4\pi^2}\int^t_{t_0}dt^\prime y_t^2}-1\right]  + \cdots  ,  
\eea
where ``$\cdots$'' does not contain $m_0^2$ and $A_0$. 
As emphasized in \eq{key}, 
the most important notice should be taken here that 
$e^{\frac{3}{4\pi^2}\int^t_{t_0}dt^\prime y_t^2}\approx \frac13$
for $t=t_0+{\rm log}\frac{10^{2}~{\rm GeV}}{M_G}$ ($\equiv t_W$) when $\tb$ is moderately small \cite{FMM1}. 
Thus, if a universal soft squared mass is assumed, $m_{h_u0}^2=m_{u^c_30}^2=m_{q_30}^2\equiv m_0^2$, and $A_0=0$ is set at the GUT scale,  
Eqs.~(\ref{appxH})--(\ref{appxQ}) are recast into \cite{FMM1}
\bea \label{FP0}
&&m_{h_u}^2(t_W)
\approx\frac{3m_0^2}{2}\left[e^{\frac{-3}{4\pi^2}\int^{t_0}_{t_W}dt^\prime y_t^2}-\frac13\right] +\cdots ~\approx~ 0.006 ~m_0^2 ~+~ \cdots ,
\\
&&m_{u^c_3}^2(t_W)\approx \frac{3m_0^2}{2}\left[\frac23 ~e^{\frac{-3}{4\pi^2}\int^{t_0}_{t_W}dt^\prime y_t^2}+0\right] +\cdots ~\approx~  \frac13 m_0^2 ~+~ \cdots ,
\\
&&m_{q_3}^2(t_W)\approx \frac{3m_0^2}{2}\left[\frac13 ~e^{\frac{-3}{4\pi^2}\int^{t_0}_{t_W}dt^\prime y_t^2}+\frac13\right] +\cdots ~\approx~ \frac{2}{3}m_0^2 ~+~ \cdots ,
\eea
where ``$\cdots$'' does not contain $m_0^2$. 
Hence, $m_{h_u}^2(t)$ almost vanishes at the EW sale ($t\approx t_W$).  
%
%
It means that $m_{h_u}^2$ can be light enough at the EW scale, almost {\it independent of $m_0^2$}, only if the ``$\cdots$'' in \eq{FP0} was also suppressed. 
Since $m_{h_u}^2$ is very insensitive to $m_0^2$, 
even a large enough $m_0^2$ guarantees the smallness of $m_{h_u}^2$ at the EW scale, 
whereas it makes stop masses quite heavy: $m_{u^c_3}^2(t_W)\approx m_0^2/3$ and $m_{q_3}^2(t_W)\approx 2m_0^2/3$.  
In the FP scenario, therefore, the naturalness of the EW scale and the Higgs mass is based on Natural tuning.

%
Although $A_0$ is comparable to other soft parameters, $m_{h_u}^2$ can still remain small  
at the EW scale, provided $(m_{h_u0}^2, m_{u^c_30}^2, m_{q_30}^2, A_0^2)$ are very specially related, satisfying, e.g., $m_0^2~ (1, ~1+x-3y, ~1-x, ~9y)$ at the GUT scale, 
where $x$, $y$ are arbitrary numbers \cite{FS}.
%
However, such a relation looks hard to realize in a supergravity (SUGRA) model. 
For simplicity, we will assume in this paper that $|x|, |y|\ll 1$; namely, 
$A_0$ is quite suppressed compared to $m_0^2$ ($=m_{h_u0}^2=m_{u^c_30}^2=m_{q_30}^2$). 
Actually, this is possible, e.g., in the gauge mediated SUSY breaking scenario with a GUT scale messenger. 
To get a universal soft squared mass in the gauge mediation, 
the SM gauge group should be embedded in a simple group at the GUT scale. 
However, the effect by nonvanishing $A_0$ on $m_{h_u}^2$ can be compensated 
by another ingredient introduced later. 
Hence, the gravity mediated SUSY breaking 
scenario with the universal soft squared mass and $A_0\neq 0$ can also be 
consistent with the FP scenario.

Unlike the naive expectation, the low energy value of $m_{h_u}^2$ is not sensitive to the stop masses in the FP scenario. 
Hence, apparently, the naturalness of the Higgs boson seems to be guaranteed in this framework. 
It is a result of 

{\bf 1.} the employed initial conditions,  $m_{h_u0}^2=m_{u^c_30}^2=m_{q_30}^2=m_0^2$ and $A_0=0$, and 

{\bf 2.} the accidental result, $e^{\frac{-3}{4\pi^2}\int^{t_0}_{t_W}dt^\prime y_t^2}\approx \frac13$ ~(``Natural tuning''). 

\noindent 
The first condition is associated with a model-building problem. 
Actually, it can easily be realized in a large class of simple SUGRA models. 
However, the second condition would be a kind of fine-tuning condition, 
because the top quark Yukawa coupling $y_t(t)$ and the interval of the energy scales between the EW and the GUT scales should specially be related. 
But it is not artificially designed.   
As mentioned in the introduction, we will simply accept such a Natural tuning phenomenon.

However, the recent experimental results at the LHC seem to spoil the nice picture 
of the original FP scenario. 
Most of all, the gauge contributions in Eqs.~(\ref{RGsol1})--(\ref{RGsol4}) cannot be ignored any longer, since the mass bound for the gluino has been increased, $M_3\gtrsim 1.4~{\rm TeV}$ \cite{gluinomass}. 
As a result, the unified gaugino mass $m_{1/2}$ should be heavier than at least $550~{\rm GeV}$. 
Since a large $m_{1/2}^2$ leads to a large negative $m_{h_u}^2$ and large positive $m_{u^c_3}^2$ and $m_{q_3}^2$ at low energy, as seen in Eqs.~(\ref{RGsol1})--(\ref{RGsol3}) and (\ref{F(t_T)}), 
$-m_{h_u}^2$ cannot be small enough at the EW scale.  
A too large negative $m_{h_u}^2$ should be finely tuned with $|\mu|^2$ to be matched to $M_Z^2$ in \eq{extremeCondi}.  
Moreover, the observed Higgs mass, $126 \gev$, is somewhat heavy as a SUSY Higgs mass. 
Once we suppose $A_0\approx 0$, a quite heavy stop mass ($\sim 5 \tev$) is needed for explaining the observed Higgs mass \cite{3-loop}.\footnote{To be precise, a $3$--$5 \tev$ stop mass is needed for a 126 GeV Higgs mass at three-loop level when $A_0=0$.  
According to Ref.~\cite{3-loop}, parametric uncertainty in the top quark mass ($m_t^{\rm pole}=173.3\pm 1.8 \gev$) results in uncertainty of 0.5 to 2 GeV in the Higgs mass. 
Among public codes providing the two-loop results, moreover,   
inconsistencies of up to 4 GeV are observed.  
In this paper, we adopt the three-loop result of Ref.~\cite{3-loop}. 
To be conservative, however, we will take 5 TeV as the stop mass 
needed for the 126 GeV Higgs mass, although a stop mass lighter than 5 TeV turns out to further decrease the fine-tuning.} 
A very large $m_{1/2}^2$ for a $5 \tev$ stop mass would require a serious fine-tuning between $m_{h_u}^2$ and $|\mu|^2$ or $m_{1/2}^2$ and $m_0^2$. 
Alternatively, one can try to extend the MSSM for raising the Higgs mass. 
However, many extensions of the MSSM Higgs sector end up   
ruining the FP scenario, as will be commented later. 

Since the stops are decoupled around $5~{\rm TeV}$ ($t\equiv t_T$), 
$m_{h_u}^2$ follows the RG running of the SM below $t\approx t_T$. 
Hence, the FP mechanism based on the SUSY RG equations would not work well anymore. 
Actually, \eq{FP0} is valid when  
the stop is not too much heavier than the $Z$ boson. 
The heavy fields' correction to the RG solution can be estimated using the formula on the Coleman-Weinberg's effective potential \cite{CW}.  
In fact, the RG solution is a result of one-loop effects by massless fields, 
while the Coleman-Weinberg's one-loop effective potential is dominated by the heavy fields. 
The signs of both loop effects are opposite.  
Thus, the low energy value of $m_{h_u}^2$ below the stop decoupling scale is roughly estimated as \cite{CQW,book} 
\dis{ \label{RGsm}
m_{h_u}^2(t_W)&\approx m_{h_u}^2|_{\Lambda_T} + \frac{3|y_t|^2}{8\pi^2}\left[(\widetilde{m}_t^2+m_t^2)\left\{{\rm log}\frac{\widetilde{m}_t^2+m_t^2}{\Lambda_T^2}-1\right\}-m_t^2\left\{{\rm log}\frac{m_t^2}{\Lambda_T^2}-1\right\}\right]
\\
&\approx m_{h_u}^2|_{\Lambda_T} - \frac{3|y_t|^2}{8\pi^2}\widetilde{m}_t^2 , 
}  
where $m_t$ ($\widetilde{m}_t$) denotes the top quark (stop) mass, 
and the cutoff $\Lambda_T$ [$\approx (\widetilde{m}_t^2+m_t^2)^{1/2}$] is the scale where the stops are decoupled, and so $m_{h_u}^2|_{\Lambda_T} = m_{h_u}^2(t_T)$.  
Here we set $m_{u^c_3}^2\approx m_{q_3}^2\equiv\widetilde{m}_t^2$ for simple estimation.  
Note that $\frac{3|y_t|^2}{8\pi^2}\widetilde{m}_t^2\approx (800 \gev)^2$. 
Accordingly, $m_{h_u}^2$ at $t=t_T$ (or $m_{h_u}^2|_{\Lambda_T}$) should be smaller 
than $(1~{\rm TeV})^2$ in order for $-m_{h_u}^2$ at the EW scale to be smaller 
than $(1~{\rm TeV})^2$.  
Since $t=t_T$ is more or less far from $t_W$, however, the coefficient of $m_0^2$ in \eq{RGsol1} is not suppressed enough, $m_{h_u}^2(t_T)\approx 0.1 m_0^2-\cdots$, 
where $m_0^2 > (5~{\rm TeV})^2$ for obtaining $5~{\rm TeV}$ stop masses.  
Hence, $m_{h_u}^2(t_T)$ is quite sensitive to $m_0^2$, 
and it should be tuned with $m_{1/2}^2$ in \eq{RGsol1} and/or $|\mu|^2$.  
Thus, for a predictively small $m_{h_u}^2$, the FP should somehow appear around the stop decoupling  scale \cite{M3/M2model,DQW}. 
That is to say, the coefficient of $m_0^2$ should be much closer to zero around the stop mass scale, as mentioned in the introduction.

\begin{figure}
\begin{center}
\subfigure[]
{\includegraphics[width=0.48\linewidth]{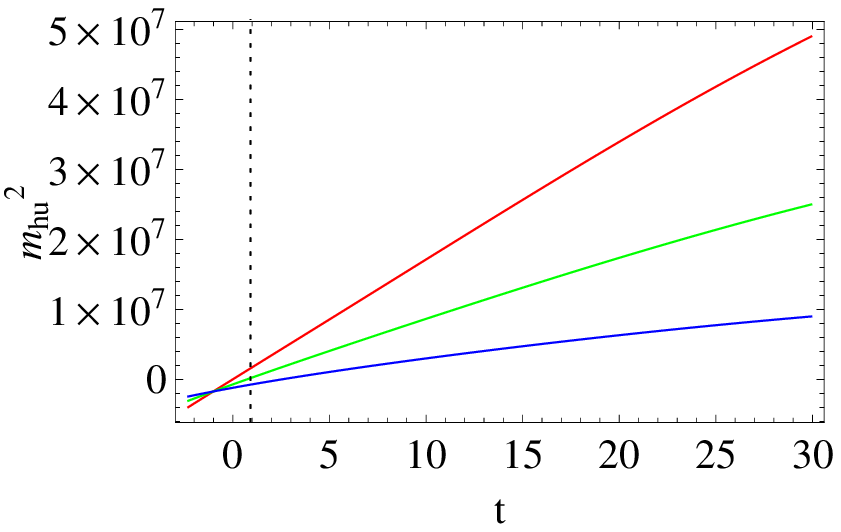}}
\hspace{0.2cm}
\subfigure[] 
{\includegraphics[width=0.48\linewidth]{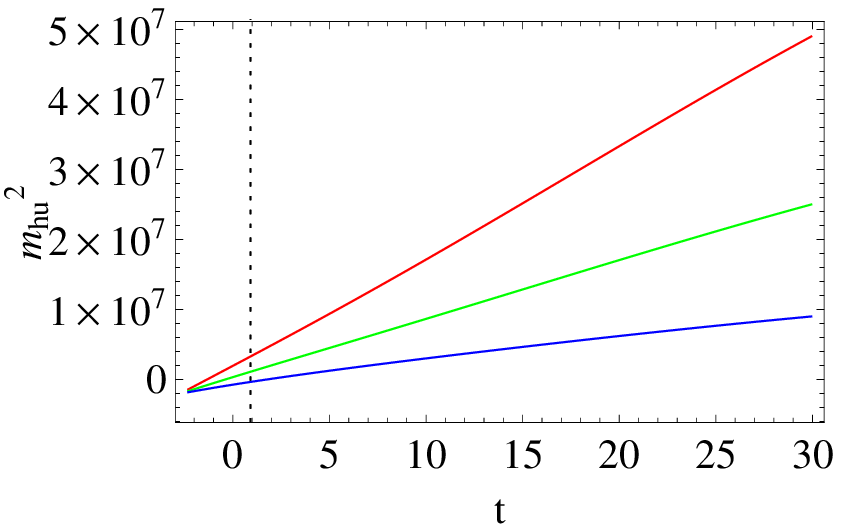}}
\end{center}
\caption{RG evolutions of $m_{h_u}^2$ for $m_0^2=(7 \tev)^2$ (red), 
$(5 \tev)^2$ (green), and $(3 \tev)^2$ (blue), and for {\bf (a)} $\tb=5$  and 
{\bf (b)} $\tb=50$, when $m_{1/2}=1 \tev$ and $A_0=0$. Here we take $\alpha_{G}=1/25$. 
The unit of the vertical axis is $({\rm GeV})^2$. 
The dotted lines at $t\approx 0.92$ denote the assumed stop decoupling scale, $Q=5 \tev$.  
$t\approx -2.3$ [$t\approx 29.9$] corresponds to $Q=200 \gev$ [$Q=2\times 10^{16} \gev$]. 
Below the stop decoupling scale, the above RG runnings must be modified. 
The above figures show that the extrapolated FP, where $m_{h_u}^2$ is negative, appears at a relatively higher (lower) 
energy scale for small (large) $\tb$.  
} \label{fig:MSSM}
\end{figure}

Figs.~\ref{fig:MSSM}-(a) and (b) display the RG behaviors of $m_{h_u}^2$ for $m_0^2=(7 \tev)^2$, $(5 \tev)^2$, $(3 \tev)^2$, when $m_{1/2}=1 \tev$, $A_0=0$, $\tb=5$ [Fig.~\ref{fig:MSSM}-(a)] and $\tb=50$ [Fig.~\ref{fig:MSSM}-(b)] with $\alpha_{G}=1/25$. 
Note that $m_{1/2}=1~{\rm TeV}$ yields the gluino mass of $2.4~{\rm TeV}$ at TeV scale, 
which is well above the present experimental lower bound $1.4 \tev$ \cite{gluinomass}.
Although we presented the simple RG equations valid for small $\tb$ in Eqs.~(\ref{RG1})--(\ref{RG3}), 
the figures in Fig.~\ref{fig:MSSM} are based on the full one-loop RG equations 
including $y_b$ and $y_\tau$ with the universal boundary condition imposed also for $m_{h_d}^2$, $m_{e^c_3}^2$, and $m_{l_3}^2$.  
Figs.~\ref{fig:MSSM}-(a) and (b) show that 
the FP is located at a slightly higher (lower) energy scale for a small (large) $\tb$.  
Table~\ref{tab:FP0} lists the values of $\{m_{q_3}^2,m_{u^c_3}^2,m_{h_u}^2\}$ at $t=t_T$ (i.e. at $Q=5~{\rm TeV}$) in these cases.  
It shows that $m_{h_u}^2(t_T)$ is quite sensitive to $m_0^2$, 
as mentioned above. 
For $\tb=50$, particularly, the fine-tuning measure defined in Refs.~\cite{FTmeasure} 
is estimated as 
\dis{ \label{FTmeasure0}
\Delta_{m_0^2}=\left|\frac{\partial ~{\rm log} ~m_Z^2}{\partial ~{\rm log} ~m_0^2}\right|
=\left|\frac{m_0^2}{m_Z^2} ~\frac{\partial m_Z^2}{\partial m_0^2}\right|
~\approx~ 875 
}
around the $m_0^2=(7 \tev)^2$. A similar analysis with $\alpha_{G}=1/24$ turns out to yield a worse result, $\Delta_{m_0^2}\approx 1474$. They are quite large. 
It is because the locations of their FPs are too far from the point $(t=t_T, m_{h_u}^2=0)$.

%
%
%
\begin{table}[!h]
\begin{center}
\begin{tabular}
{c|ccc||c|ccc}
 $$ & $$  & {\small $\tb=5$}  & $$  & $$  & $$  & {\small $\tb=50$}  & $$
\\ \hline
 {\footnotesize ${\bf m_0^2}$} & {\footnotesize $({\bf 7} \tev)^2$}  & {\footnotesize $({\bf 5} \tev)^2$}  & {\footnotesize $({\bf 3} \tev)^2$}  & 
 {\footnotesize ${\bf m_0^2}$}  & {\footnotesize $({\bf 7} \tev)^2$}  & {\footnotesize $({\bf 5} \tev)^2$}  & {\footnotesize $({\bf 3} \tev)^2$}
\\ \hline
 {\footnotesize $m_{q_3}^2(t_T)$} & {\footnotesize $(6.1 \tev)^2$}  & {\footnotesize $(4.5 \tev)^2$}  & {\footnotesize $(3.1 \tev)^2$}  & 
 {\footnotesize $m_{q_3}^2(t_T)$}  & {\footnotesize $(5.2 \tev)^2$}  & {\footnotesize $(3.9 \tev)^2$}  & {\footnotesize $(2.8 \tev)^2$}
\\
 {\footnotesize $m_{u^c_3}^2(t_T)$} & {\footnotesize $(4.6 \tev)^2$}  & {\footnotesize $(3.4 \tev)^2$}  & {\footnotesize $(2.4 \tev)^2$}  & {\footnotesize $m_{u^c_3}^2(t_T)$} & {\footnotesize $(4.7 \tev)^2$}  & {\footnotesize $(3.5 \tev)^2$}  & {\footnotesize $(2.5 \tev)^2$}
\\
 {\footnotesize ${\bf m_{h_u}^2(t_T)}$} & {\footnotesize $({\bf 1.3} \tev)^2$}  & {\footnotesize $-({\bf 0.4} \tev)^2$}  & {\footnotesize $-({\bf 0.9} \tev)^2$}  & {\footnotesize ${\bf m_{h_u}^2(t_T)}$}  & {\footnotesize $({\bf 1.8} \tev)^2$}  & {\footnotesize $({\bf 1.1} \tev)^2$}  & {\footnotesize $-({\bf 0.6} \tev)^2$}
\end{tabular}
\end{center}\caption{Soft squared masses of the stops and Higgs boson at $Q=5 \tev$ for 
$m_0^2=(7 \tev)^2$, $(5 \tev)^2$, and $(3 \tev)^2$,  
when $m_{1/2}=1 \tev$ and $A_0=0$ with $\alpha_{G}=1/25$. The left (right) four columns correspond to the results of $\tb =5$ ($\tb =50$). 
}\label{tab:FP0}
\end{table}
%
%

%

\begin{figure}
\begin{center}
%
{\includegraphics[width=0.70\linewidth]{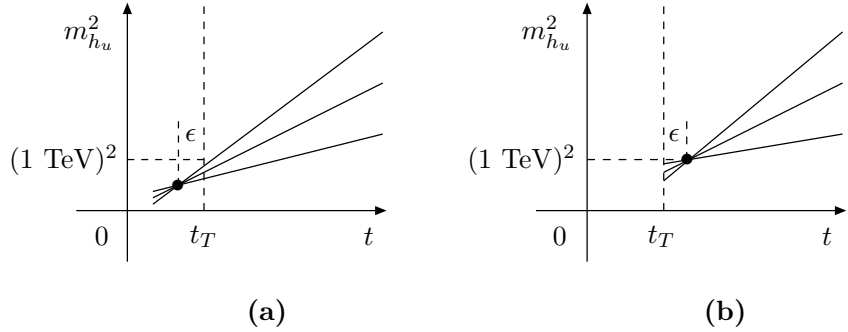}}
\end{center}
\caption{Desirable locations of the focus point in the $(t, m_{h_u}^2)$ space.  
The straight lines sketch different RG evolutions of $m_{h_u}^2$ for various $m_0^2$s. 
$t_T$ corresponds to the assumed stop decoupling scale ($Q=5 \tev$). 
$\epsilon$ needs to be as small as possible.  
}\label{fig:desirable}
\end{figure}

In order to get $m_{h_u}^2$ that is small enough and insensitive to $m_0^2$, 
the location of the FP needs to be moved somehow to a position {\it around} the stop mass scale.  
See Fig.~\ref{fig:desirable}.    
$\epsilon$ in Figs.~\ref{fig:desirable}-(a) and (b) should be as small as possible
for a predictable $m_{h_u}^2$ at the EW scale. 
In addition, at a location of the FP near $t=t_T$,  
$m_{h_u}^2$ should be in the range of $0\lesssim m_{h_u}^2\lesssim (1 \tev)^2$.
Since the heavy gluino makes a large negative contribution to $m_{h_u}^2(t_T)$, 
we need some other ingredients to overcome the heavy gluino effect. 
Below $t=t_T$, $m_{h_u}^2$ further decreases by $\sim (800 \gev)^2$ down to $t=t_W$, 
as discussed in \eq{RGsm}.
In order to mitigate the $m_0^2$ dependence via $\widetilde{m}_t^2$ in \eq{RGsm}, 
reducing the fine-tuning, 
a FP of $m_{h_u}^2$ appearing at a slightly lower energy scale than (but still around) $t_T$ is more preferred: the coefficient of $m_0^2$ in $m_{h_u}^2|_{\Lambda_T}$ needs to be of order ${\cal O}(10^{-2})$. 
%


%
%


\section{Precise focusing}  \label{sec:preciseFP}


In this section, we will discuss how to move the FP to the desirable locations presented in the previous section in the $(t,m_{h_u}^2(t))$ space.  
We intend to argue that the Higgs mass happens to be 126 GeV by $5 \tev$ stop mass,
after $m_{h_u}^2$ at $t=t_T$ is made insensitive to $m_0^2$. 
It would be a way to trim the original idea of the Natural tuning.

\subsection{Pushing up the focus point to higher energy scale}

As $\tb$ increases, the size of the top quark Yukawa coupling decreases. 
As a consequence, the factor $[e^{\frac{-3}{4\pi^2}\int^{t_0}_{t}dt^\prime y_t^2}-\frac13]$ 
in \eq{FP0} vanishes at a lower energy scale $t$ ($< t_W$) for a smaller $y_t$. 
It implies that the FP moves to a lower energy scale for a larger $\tb$ \cite{FMM1,Akula}. 
The numerical analysis including $y_b$ and $y_\tau$, 
Figs.~\ref{fig:MSSM}-(a) and (b) confirm such a behavior of the FP.  
Since we intend to move the FP in the higher energy direction, 
a large $\tb$ is not helpful.

A much larger top quark Yukawa coupling $y_t(t)$ at {\it higher} energy scales can move the FP 
to a new location at a higher energy scale.  
Actually, $y_t(t)$ can be easily raised at higher energy scales 
e.g. by introducing a new Yukawa coupling of the Higgs boson. 
For instance, let us consider a coupling between $h_u$ and a new singlet $S$ in the next-to-MSSM (NMSSM) \cite{NMSSMreview}: 
\dis{
W_S=\lambda Sh_uh_d + \cdots .
} 
In this case, the RG equations of $y_t$ and $\lambda$ are given by 
\bea \label{RGnmssm}
&&8\pi^2\frac{d}{dt}y_t^2=y_t^2\left[\lambda^2+6y_t^2 
-\frac{16}{3}g_3^2-3g_2^2-\frac{13}{15}g_1^2\right] ,
\\
&&~~~ 8\pi^2\frac{d}{dt}\lambda^2=\lambda^2\left[4\lambda^2+3y_t^2
-3g_2^2-\frac35 g_1^2\right] 
\eea    
for small $\tb$. 
Because of the additional positive contribution by $\lambda^2$ 
to the RG equation of $y_t$, 
$y_t^2$ becomes larger than that in the absence of $\lambda$. 
Moreover, the $\lambda$ coupling introduces a positive contribution 
also to the RG equation for $m_{h_u}^2$:
\dis{ \label{mhuLambda}
16\pi^2\frac{d}{dt}m_{h_u}^2=2\lambda^2\left(X_\lambda + A_\lambda^2\right)
+ 6y_t^2\left(X_t+A_t^2\right)
-6g_2^2M_2^2-\frac65g_1^2M_1^2 ,
}
where $X_\lambda\equiv(m_{h_u}^2+m_{u^c_3}^2+m_{q_3}^2)$. 
It turns out, however, that the FP's location is too sensitive to $\lambda$. 
According to our analysis, 
$\lambda$ should be smaller than at least $0.1$. 
Otherwise, the FP moves too far away in the high energy direction. 
For example, $\lambda=0.6$ and $\tb=3$ moves the location of FP to $10^{13} \gev$ energy scale.
Hence, the parameter window satisfying the $126 \gev$ Higgs mass and the Landau pole constraint in the NMSSM, $0.6\lesssim\lambda\lesssim 0.7$ and $1<\tb\lesssim 3$ \cite{nmssmWindow}, cannot be compatible with the FP scenario.  
As seen in this example, extensions of the MSSM Higgs sector with a new sizable Yukawa coupling,
e.g., for raising the Higgs mass could result in ruin of the FP scenario.\footnote{With a relatively lighter stop mass ($\lesssim 1 \tev$), the (singlet) extensions of the MSSM can significantly reduce the fine-tuning 
by adding an additional tree level \cite{NMSSMreview,singletEXT} or a radiative Higgs mass \cite{extensions}.}

The RG effect of $\lambda$ coupling on $y_t$ can be reduced just by 
assuming that $S$ is superheavy and so decoupled at a very high energy scale. 
One well-motivated superheavy particle is the RH neutrino ($N^c$), 
which is introduced to explain the smallness of the active neutrino mass 
through the seesaw mechanism \cite{seesaw} by the superpotential,  
%
%
\dis{ \label{RHnu}
W_N=y_Nl_3h_uN^c + \frac12M_NN^cN^c ,  
}
where $l_3$ is a lepton doublet in the MSSM. 
We assume that the Majorana mass of $N^c$ is $M_N\approx 2\times 10^{14} \gev$. 
If the RH neutrino is embedded in a multiplet of a GUT with the $B-L$ charge, 
\eq{RHnu} can be naturally obtained from the nonrenormalizable term in GUTs, 
$W\supset \langle H_G\rangle\langle H_G\rangle N^cN^c/M_P$, 
where $\langle H_G\rangle$ 
and $M_P$ are a VEV of a GUT breaking Higgs boson ($\sim 10^{16} \gev$) and 
the reduced Planck mass ($\approx 2.4\times 10^{18} \gev$), respectively.    
For $M_N\sim 10^{14}~{\rm GeV}$, the Yukawa coupling $y_N$ should be of order unity to get a neutrino mass of order $0.1 \ev$. 
Here, we suppose that only one Yukawa coupling with $h_u$, $y_N$ is of order unity:   
for simplicity, we assume that other Yukawa couplings of $h_u$ to other RH neutrinos 
are small enough. Accordingly, other RH neutrinos should be relatively lighter than $M_N$. 
%
Since $N^c$ would be decoupled at a very hight energy scale 
($Q= M_N\approx 2\times 10^{14} \gev$), 
its RG effect on $y_t$ could be mild, and the FP would relatively slowly move as $y_N$ varied. 
Consequently, $m_{h_u}^2$ at $t=t_T$ could become less sensitive to $m_0^2$ \cite{RHnuFP}. 
If the heaviest RH neutrino was lighter than $\sim 10^{13} \gev$, its RG effect on $y_t$ would be negligible because the required Yukawa coupling becomes too small.

Similar to \eq{mhuLambda}, the RG evolution of $m_{h_u}^2$ between $Q=M_G$ and $Q= M_N$ is described by   
\begin{eqnarray} \label{mHu2I}
16\pi^2\frac{d}{dt}m_{h_u}^2&=&2y_N^2\left(X_N+A_N^2\right)+6y_t^2\left(X_t+A_t^2\right)
-6g_2^2M_2^2-\frac65 g_1^2M_1^2 , 
\end{eqnarray}
where the $y_N^2X_N$ [$=y_N^2(m_{h_u}^2+m_{N^c}^2+m_{l_3}^2)$] and $y_N^2A_N^2$ terms 
are additional positive contributions coming from the RH neutrino. 
On the other hand, the RG equations for $m_{u^c_3}^2$ and $m_{q_3}^2$ 
maintain the same forms with those 
in the absence of the RH neutrino, Eqs.~(\ref{RG2}) and (\ref{RG3}). 
They are just affected only through the modified value of $y_t^2\left(X_t+A_t^2\right)$, 
which appears also in \eq{mHu2I}. 
For the complete form of the RG equations, refer to the Appendix. 
Because of the $y_N^2\left(X_N+A_N^2\right)$ terms in \eq{mHu2I}, 
$m_{h_u}^2/m_{u^c_3}^2$ and $m_{h_u}^2/m_{q_3}^2$ more rapidly decrease 
from $Q=M_G$ to $Q= M_N$ than the case without the RH neutrino. 
Below $Q= M_N$, however, the RH neutrino becomes decoupled, 
and so $m_{h_u}^2$, $m_{u^c_3}^2$, and $m_{q_3}^2$ respect the same RG equations 
with Eqs.~(\ref{RG1})--(\ref{RG3}). 

Considering \eq{RGsol1}, one can see that the RG solution of $m_{h_u}^2$ {\it valid only below} $Q= M_N$ ($t<t_I$) should be written as 
\dis{ \label{solI}
&\qquad~~~  m_{h_u}^2(t) = m_{h_uI}^2+\frac{X_I}{2}
\left[e^{\frac{3}{4\pi^2}\int^{t}_{t_I}dt^\prime y_t^2}-1\right] 
+ \cdots 
\\
&=\frac{X_I}{2}\left[e^{\frac{-3}{4\pi^2}\int^{t_I}_{t}dt^\prime y_t^2}-\left(1-\frac{2m_{h_uI}^2}{m_{h_uI}^2+m_{u^c_3I}^2+m_{q_3I}^2}\right)\right]+\cdots , 
}
where $\{m_{h_uI}^2, m_{u^c_3I}^2, m_{q_3I}^2\}$ denote the values of $\{m_{h_u}^2, m_{u^c_3}^2, m_{q_3}^2\}$ at $Q= M_N$, respectively, and $X_I\equiv m_{h_uI}^2+m_{u^c_3I}^2+m_{q_3I}^2$. 
Note that ``$\cdots$'' in \eq{solI} does not contain the dependence of $\{m_{h_uI}^2, m_{u^c_3I}^2, m_{q_3I}^2\}$. 
Comparing with \eq{RGsol1}, $\{m_{h_u0}^2, m_{u^c_30}^2, m_{q_30}^2\}$ and $X_0$ are replaced 
by $\{m_{h_uI}^2, m_{u^c_3I}^2, m_{q_3I}^2\}$ and $X_I$ in \eq{solI}. 
On the contrary, $y_t^2$ in \eq{solI} is the same as $y_t^2$ of \eq{RGsol1} for $t<t_I$, 
because $y_t^2$ should be set to explain the top quark mass at low energy and undergoes the same RG evolution as the case of \eq{RGsol1}. 
The RH neutrino makes $y_t^2$ larger only above $Q= M_N$. 
Since $m_{h_uI}^2/m_{u^c_3I}^2$ and $m_{h_uI}^2/m_{q_3I}^2$ are more suppressed at $Q= M_N$ by the RH neutrino effect above $Q= M_N$, 
$1-2m_{h_uI}^2/(m_{h_uI}^2+m_{u^c_3I}^2+m_{q_3I}^2)$ or $1-2m_{h_uI}^2/X_I$ in \eq{solI} is larger 
than that evaluated at $Q= M_N$ in the absence of the RH neutrino. 
As a result, ${\rm exp}[\frac{-3}{4\pi^2}\int^{t_I}_{t}dt^\prime y_t^2]-(1-2m_{h_uI}^2/X_I)$ vanishes at a $t$ larger than $t_W$.
It implies that {\it a FP must still exist and appear at a scale higher than $t_W$}. 
Therefore, we can move the FP to around $t=t_T$ using a sizable $y_N$.
We will discuss it again later.

\subsection{Uplifting the focus point}

Toward the desirable FP location, we need to somehow lift up the FP in the $(t,m_{h_u}^2(t))$ space as mentioned before.   
As a trial, let us turn on a small $A_0$ in \eq{RGsol4}, 
keeping $m_{h_u0}^2=m_{u^c_30}^2=m_{q_30}^2=m_0^2$. 
Then \eq{appxH} yields $m_{h_u}^2(t_W)\approx -A_0^2/9$. 
So the FP moves in the opposite direction to our desire. 
From Eqs.~(\ref{RGsol1}) and (\ref{F(t_T)}), 
increase of $m_{1/2}^2$ also moves the FP in the negative direction. 
Because of the experimental gluino mass constraint ($M_3\gtrsim 1.4~{\rm TeV}$), 
however, one cannot decrease $m_{1/2}^2$ sufficiently.  

Indeed, the largest negative contribution to $m_{h_u}^2$ comes from the gluino mass $M_3$, 
as seen from Eqs.~(\ref{F})--(\ref{F(t_T)}): 
\eq{F} is dominated by the $g_3^2M_3$ and $g_3^2M_3^2$ terms in Eqs.~(\ref{GA}) and (\ref{GX2}), 
which eventually give a negative $F(t_T)$ as seen in \eq{F(t_T)}. 
A too large negative $m_{h_u}^2$ at the EW scale 
should be fine-tuned with $|\mu|^2$ to yield the desired size of $m_Z^2$.
One way to compensate the negative gluino mass effect on $m_{h_u}^2$ 
is to cancel it with the positive contribution from the wino mass effect, 
sacrificing the gaugino mass unification, 
$M_3^2\lesssim M_2^2$ at the GUT scale \cite{M3/M2,M3/M2model}:
such nonuniversal gaugino masses at the GUT scale      
could improve the FP behavior but also soften significantly the limits on the gluino mass.
%
Alternatively, a fine-tuning between $m_0^2$ and $m_{1/2}^2$ could also leave a light enough $m_{h_u}^2$, as seen in Eqs.~(\ref{RGsol1}) and (\ref{F(t_T)}): 
a FP achieved through such a fine-tuning 
can remain insensitive e.g. to the scaling of $(m_0^2,m_{1/2}^2)\rightarrow \lambda^2(m_0^2,m_{1/2}^2)$, 
keeping the ratio between $m_0^2$ and $m_{1/2}^2$ \cite{DQW}. 
However, the idea of Natural tuning is lost in this mechanism. 

In this paper, we propose to consider the two-loop gauge effects by  
the first and second generations of hierarchically heavier sfermions, maintaining the gaugino mass unification. 
Their two-loop Yukawa interactions are extremely suppressed by their tiny Yukawa couplings. 
For simplicity, we suppose a universal heavy mass for them ($\equiv\widetilde{m}^2$). 
If $\widetilde{m}^2\gg m_{1/2}^2$, the RG running of $\widetilde{m}^2$ is negligible. 
Then the gauge contributions to the RG equations for the soft masses of the Higgs boson and sfermions are modified as \cite{2-loop,Natural/2-loop}
\dis{ \label{1-2loops} 
&16\pi^2\frac{d}{dt}m_f^2=-8\sum_{i=3,2,1}C^f_i\left(g_i^2M_i^2
-\frac{\widetilde{m}^2}{4\pi^2}g_i^4\right) + \cdots    
\\
&~~ =-8\sum_{i=3,2,1}C^f_i\left[\left(\frac{m_{1/2}}{g_0^2}\right)^2g_i^6
-\frac{\widetilde{m}^2}{4\pi^2}g_i^4\right] + \cdots ,
}
where $f=h_u,~u^c_3,~q_3$, etc., and $C_i^f$ denotes the Casimir for $f$. 
With the universal soft mass condition, the contributions by the ``$D$-term'' potential to \eq{1-2loops} vanish. 
Since  $g_i^2M_i^2$s are always accompanied with $-\frac{\widetilde{m}^2}{4\pi^2}g_i^4$ 
in Eqs.~(\ref{RG1})--(\ref{RG3}), they all should be modified into $g_i^2M_i^2-\frac{\widetilde{m}^2}{4\pi^2}g_i^4$. 
As a result, the heavy gluino effect can be compensated to be milder by the $\widetilde{m}^2$ terms \cite{Natural/splitZprime}. 
If $\widetilde{m}$ is much heavier than the gluino mass, moreover, 
it can be comparable to it or even dominate over it. 
Thus, a heavy enough $\widetilde{m}^2$ could raise $m_{h_u}^2$ up even to a positive value at $t=t_T$. 
Note that $\widetilde{m}^2$ does not appear in $X_0$ in \eq{RGsol1}: 
the heavier sfermions' effects on Eqs.~(\ref{RG1})--(\ref{RG4}) 
via the Yukawa interactions are extremely tiny. 
So $\widetilde{m}^2$ does not touch the FP mechanism. 
Indeed, any Yukawa couplings and $\tb$ are not involved in $g_i^2M_i^2-\frac{\widetilde{m}^2}{4\pi^2}g_i^4$.     
Since both contributions originate from the gauge interactions,   
their relation could be more easily realized in a UV model \cite{ongoing} 
than the relation between $m_{1/2}^2$ and $m_0^2$. 
Note that they leave intact the $A$-term RG equation \eq{RG4}. 
For the full expressions of the semianalytic solutions, refer to the Appendix.

The hierarchical mass pattern between the first/second and the third generations can be 
realized by employing the two different SUSY breaking mediations, e.g. the gravity or gauge mediation and U(1)$^\prime$ mediation. 
For instance, the first two generations of matter could carry nonzero (but opposite) U(1)$^\prime$ charges
and they could receive additional U(1)$^\prime$ SUSY breaking mediation effects proportional to their charge squareds \cite{Zprime} for their hierarchically heavier masses \cite{splitZprime,Natural/splitZprime}.
Their desired relation could be achieved from the hierarchy between $g_0$ and the  U(1)$^\prime$ gauge coupling, and also the messengers' masses 
with a common SUSY breaking source. 
In such a setup, a relation between $\widetilde{m}^2$ and $m_{1/2}^2$ could also be obtained. 
Since the third generation of sfermions do not carry U(1)$^\prime$ charges, 
its soft masses are determined only by the gravity mediation effect.  
$A_0$ can also remain small enough 
to avoid unwanted color breaking minimum at low energies \cite{CCB}. 
We will propose a simple model realizing a desired relation between them later. 

To summarize our discussion so far, in Table~\ref{tab:FPmove} 
we present the FP's movements for the various variations of parameters.  
We can move the FP into the desirable positions of Fig.~\ref{fig:desirable} 
by using e.g. $y_N$ and $\widetilde{m}^2$.

%
%
%
\begin{table}[!h]
\begin{center}
\begin{tabular}
{c||c|c|c|c|c}
Variations  &   
 ~~${\rm tan}\beta \Uparrow$~~ &  ~$y_t^2$, $\lambda^2$, $y_N^2 \Uparrow$~  &  ~~$A_0^2 \Uparrow$~~  
 &  ~$m_{1/2}^2 \Uparrow$~  & ~~$\widetilde{m}^2 \Uparrow$~~ 
  \\
\hline
Focus point & $\Leftarrow$ & $\Rightarrow$ & $\Downarrow$ 
& $\Downarrow$ & $\Uparrow$  
\end{tabular}
\end{center}\caption{Movement of the focus point   
for increases of the various parameters in the $(t,m_{h_u}^2(t))$ space
}\label{tab:FPmove}
\end{table}
%
%

\subsection{Numerical results}

\begin{figure}
\begin{center}
\subfigure[]
{\includegraphics[width=0.48\linewidth]{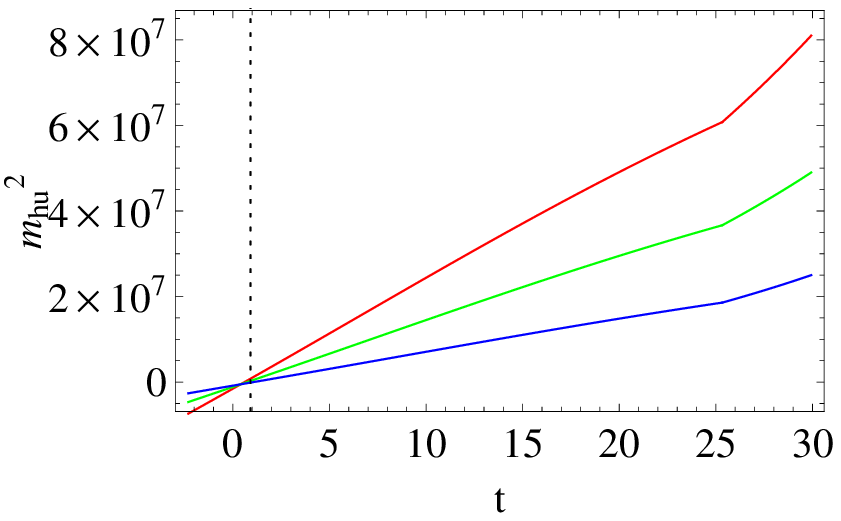}}
\hspace{0.2cm}
\subfigure[] 
{\includegraphics[width=0.48\linewidth]{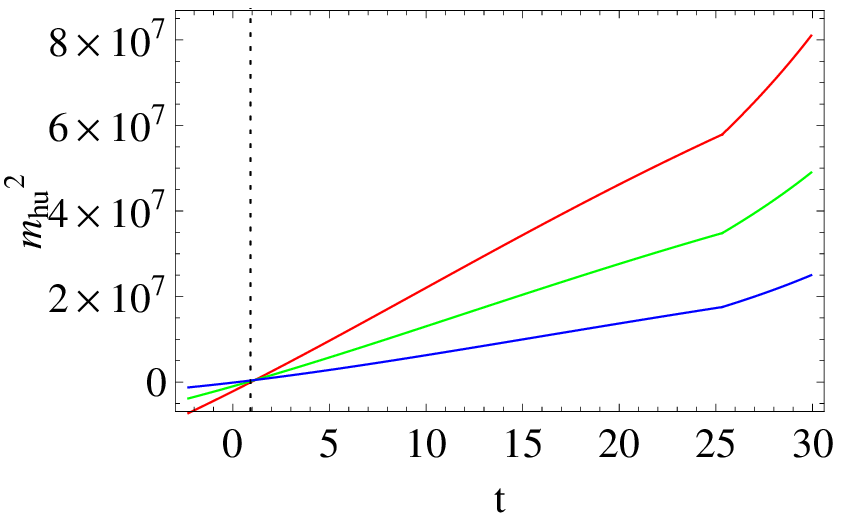}}
\end{center}
\caption{RG evolutions of $m_{h_u}^2$ for $m_0^2=(9 \tev)^2$ (red), 
$(7 \tev)^2$ (green), and $(5 \tev)^2$ (blue), and for {\bf (a)} $y_{NI}^2=0.8$, $\widetilde{m}^2=(15 \tev)^2$  and 
{\bf (b)} $y_{NI}^2=1.0$, $\widetilde{m}^2=(20 \tev)^2$, when $\tb=5$ and $m_{1/2}= A_0=1 \tev$ with $\alpha_{G}=1/24$. 
The unit of the vertical axis is $({\rm GeV})^2$. 
Below the seesaw scale, $t=t_I\approx 25.3$ [$Q\approx 2\times 10^{14} \gev$], the RH neutrino is decoupled.
The dotted lines at $t\approx 0.92$ denote the assumed stop decoupling scale, $Q=5 \tev$.  
%
Below the stop decoupling scale, the above RG runnings must be modified. 
The above figures show that the (extrapolated) FP appears at desirable locations. 
} \label{fig:FPn_5}
\end{figure}

Let us attempt to reduce the fine-tuning by introducing a superheavy RH neutrino and 
taking heavy soft masses for the first two generations of sfermions.  
Figs.~\ref{fig:FPn_5}-(a) and (b) show the numerical results for the RG evolutions of $m_{h_u}^2$ for $m_0^2=(9 \tev)^2$, $(7 \tev)^2$, and $(5 \tev)^2$, 
when $\{y_{NI}^2=0.8,~\widetilde{m}^2=(15 \tev)^2\}$  
and $\{y_{NI}^2=1.0,~\widetilde{m}^2=(20 \tev)^2\}$, respectively. 
Here, $y_{NI}$ means $y_N$ evaluated  
at the RH neutrino decoupling scale ($Q= M_N\approx 2\times 10^{14} \gev$). 
$y_{N}^2$ of $y_{NI}^2=0.8$ ($1.0$) reaches $0.95$ ($1.2$) at the GUT scale, 
while its RG evolution becomes frozen below $Q= M_N$.  
In both cases, we set $\tb=5$ and $m_{1/2}=A_0=1 \tev$ with $\alpha_{G}=1/24$. 
Note that $m_{1/2}$ and $A_0$ are U(1)$_R$ breaking parameters. 
Thus, e.g. if U(1)$_R$ breaking scale is relatively lower than the SUSY breaking scale, 
they can be smaller than other soft SUSY breaking parameters, 
$m_0^2$ and $\widetilde{m}^2$ as desired.  
In Ref.~\cite{Li}, conformal sequestering was considered to suppress them.     
In ``pure gravity mediation,'' $m_{1/2}$ and $A_0$ are suppressed at the tree level \cite{puregravity}. 
Below the seesaw scale, $t=t_I\approx 25.3$ [$Q\approx 2\times 10^{14} \gev$], the RH neutrino is decoupled.
Thus, $m_{h_u}^2$s in Figs.~\ref{fig:FPn_5}-(a) and (b) follow the RG equations without the RH neutrino below $t=t_I$, 
while they are governed 
by the full RG equations including the RH neutrino between $t=t_0$ and $t=t_I$. 
For the analyses in Figs.~\ref{fig:FPn_5}-(a) and (b), we used the full RG equations in the Appendix with the boundary conditions, 
$m_{h_u}^2=m_{u^c_3}^2=\cdots=m_{h_d}^2=\cdots=m_{N^c}^2=m_0^2$ and $m_{u^c_{1,2}}^2=m_{q_{1,2}}^2 \cdots =\widetilde{m}^2$.

 %
%
%
\begin{table}[!h]
\begin{center}
\begin{tabular}
{c|ccc||c|ccc}
 $$ & {\footnotesize $\tb=5$}  & {\footnotesize $y_{NI}^2=0.8$}~  & {\footnotesize $\widetilde{m}=15 \tev$}  & $$  & {\footnotesize $\tb=5$}  & {\footnotesize $y_{NI}^2=1.0$}  & {\footnotesize $\widetilde{m}=20 \tev$}
\\ \hline
 {\footnotesize ${\bf m_0^2}$} & {\footnotesize $({\bf 9} \tev)^2$}  & {\footnotesize $({\bf 7} \tev)^2$}  & {\footnotesize $({\bf 5} \tev)^2$}  & 
 {\footnotesize ${\bf m_0^2}$}  & {\footnotesize $({\bf 9} \tev)^2$}  & {\footnotesize $({\bf 7} \tev)^2$}  & {\footnotesize $({\bf 5} \tev)^2$}
\\ \hline
 {\footnotesize $m_{q_3}^2(t_T)$} &  {\footnotesize $(7.3 \tev)^2$}  & {\footnotesize $(5.6 \tev)^2$}  & {\footnotesize $(3.7 \tev)^2$} &
 {\footnotesize $m_{q_3}^2(t_T)$}  & {\footnotesize $(6.9 \tev)^2$}  & {\footnotesize $(5.0 \tev)^2$}  & {\footnotesize $(2.8 \tev)^2$}  
\\
 {\footnotesize $m_{u^c_3}^2(t_T)$} &  {\footnotesize $(5.7 \tev)^2$}  & {\footnotesize $(4.3 \tev)^2$}  & {\footnotesize $(2.8 \tev)^2$}  & {\footnotesize $m_{u^c_3}^2(t_T)$}  & {\footnotesize $(5.3 \tev)^2$}  & {\footnotesize $(3.8 \tev)^2$}  & {\footnotesize $(1.9 \tev)^2$} 
\\
 {\footnotesize ${\bf m_{h_u}^2(t_T)}$} & {\footnotesize $({\bf 0.9} \tev)^2$}  & {\footnotesize $({\bf 0.5} \tev)^2$}  & {\footnotesize $-({\bf 0.3} \tev)^2$}  & {\footnotesize ${\bf m_{h_u}^2(t_T)}$}  &  {\footnotesize $-({\bf 0.2} \tev)^2$}  & {\footnotesize $({\bf 0.4} \tev)^2$}  & {\footnotesize $({\bf 0.6} \tev)^2$} 
\end{tabular}
\end{center}\caption{Soft squared masses of the stops and Higgs boson at $t=t_T\approx 0.92$ ($Q=5 \tev$) for 
$m_0^2=(9 \tev)^2$, $(7 \tev)^2$, and $(5 \tev)^2$,  
when $\tb=5$ and $m_{1/2}= A_0=1 \tev$ with $\alpha_{G}=1/24$. 
The left [right] four columns correspond to the results of $\{y_{NI}^2=0.8,~\widetilde{m}^2=(15 \tev)^2\}$ [$\{y_{NI}^2=1.0,~\widetilde{m}^2=(20 \tev)^2\}$]. 
}\label{tab:FPn_5}
\end{table}

In Fig.~\ref{fig:FPn_5}-(a) [(b)], the FP appears at a slightly lower [higher] scale 
than the stop decoupling scale ($t=t_T\approx 0.92$). 
Since $m_{h_u}^2$ is well focused in the both cases, 
$m_{h_u}^2(t_T)$ is quite insensitive to the various trial $m_0^2$s as seen in Table~\ref{tab:FPn_5}: 
for $5 \tev < m_0^2 < 9 \tev$ at the GUT scale, 
$m_{h_u}^2$ just changes from $-(0.3 \tev)^2$ [$(0.6 \tev)^2$] to $(0.9 \tev)^2$ [$-(0.2 \tev)^2$] 
at the stop decoupling scale.
Hence, for precise focusing, it is required that 
\dis{
0.8~\lesssim ~y_{NI}^2~\lesssim ~1.0  \quad {\rm and} \quad  (15 \tev)^2~\lesssim~\widetilde{m}^2~\lesssim~ (20 \tev)^2 ,
}
when $\tb=5$ and $m_{1/2}=A_0= 1 \tev$.
%
Under the situation that $m_{h_u}^2$ at $t=t_T$ is insensitive to $m_0^2$ and stop masses,  
$m_0^2$ can happen to be around $(8 \tev)^2$ at the GUT scale, 
which leads to $5 \tev$ stop masses and the $126 \gev$ Higgs mass at the EW scale. 
However, if a larger $y_{NI}^2$ is taken, e.g. $y_{NI}^2=1.4$, 
the FP emerges around $t\approx 3$ ($Q\approx 40 \tev$). 
For $\widetilde{m}^2\gtrsim (24 \tev)^2$ and $y_{NI}^2=1.0$, 
the EW symmetry breaking does not arise, 
because $m_{h_u}^2(t_T) > (1 \tev)^2$.   
Hence, the above range of $y_{N}$ and $\widetilde{m}^2$ for a desirable FP 
needs to be supported by a UV model.
Once $M_N$ is fixed by a GUT as explained above, however, 
the above range of $y_{NI}^2$ could be regarded as another Natural tuning, since $y_N^2$ can be determined by the active neutrino mass. 
The tuning issue introduced for the desired $\widetilde{m}^2$
could be converted to a model-building problem \cite{ongoing}.

\begin{figure}
\begin{center}
\subfigure[]
{\includegraphics[width=0.48\linewidth]{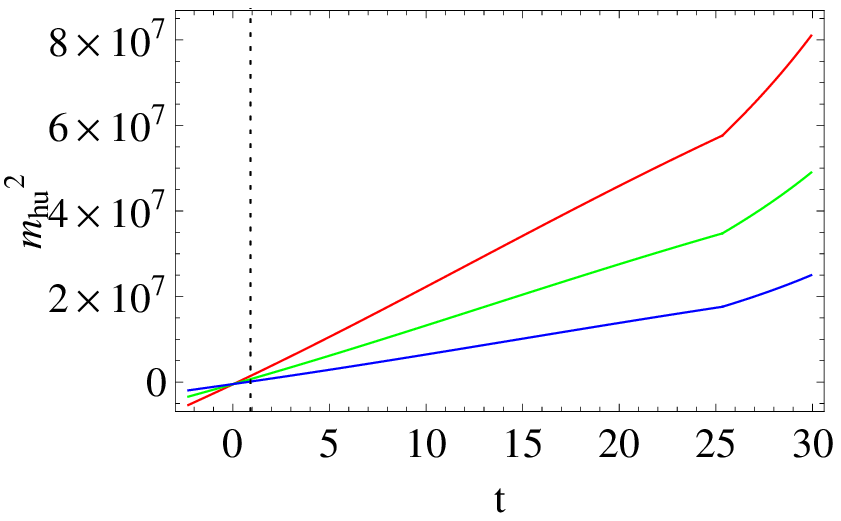}}
\hspace{0.2cm}
\subfigure[] 
{\includegraphics[width=0.48\linewidth]{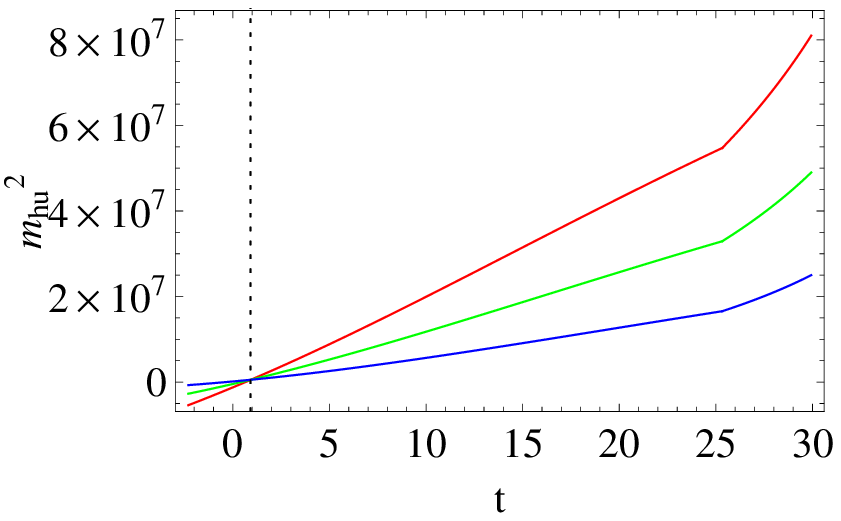}}
\end{center}
\caption{RG evolutions of $m_{h_u}^2$ for $m_0^2=(9 \tev)^2$ (red), 
$(7 \tev)^2$ (green), and $(5 \tev)^2$ (blue), and for {\bf (a)} $y_{NI}^2=1.0$, $\widetilde{m}^2=(15 \tev)^2$  and 
{\bf (b)} $y_{NI}^2=1.2$, $\widetilde{m}^2=(20 \tev)^2$, when $\tb=50$ and $m_{1/2}= A_0=1 \tev$ with $\alpha_{G}=1/24$. 
The unit of the vertical axis is $({\rm GeV})^2$. 
Below the seesaw scale, $t=t_I\approx 25.3$ [$Q\approx 2\times 10^{14} \gev$], the RH neutrino is decoupled.
The dotted lines at $t=0.92$ denote the assumed stop decoupling scale, $Q=5 \tev$.  
%
Below the stop decoupling scale, the above RG runnings must be modified. 
The above figures show that the (extrapolated) FP appears at desirable locations. 
} \label{fig:FPn_50}
\end{figure}

Similarly, Figs.~\ref{fig:FPn_50}-(a), (b), and Table~\ref{tab:FPn_50} present the results 
of $m_{h_u}^2$ for $m_0^2=(9 \tev)^2$, $(7 \tev)^2$, and $(5 \tev)^2$, when $\tb=50$ 
and $m_{1/2}= A_0=1 \tev$ with $\alpha_{G}=1/24$. 
Here, we take 
$\{y_{NI}^2=1.0,~\widetilde{m}^2=(15 \tev)^2\}$  
and $\{y_{NI}^2=1.2,~\widetilde{m}^2=(20 \tev)^2\}$ in Figs.~\ref{fig:FPn_50}-(a) and (b), respectively. 
$y_{N}^2$ of $y_{NI}^2=1.0$ ($1.2$) reaches $1.25$ ($1.6$) at the GUT scale. 
Thus, the parameter ranges required for precise focusing are 
\dis{
1.0~\lesssim ~y_{NI}^2~\lesssim ~1.2 
\quad {\rm and} \quad  (15 \tev)^2~\lesssim~\widetilde{m}^2~\lesssim~ (20 \tev)^2 , 
}
when $\tb=50$ and $m_{1/2}= A_0=1 \tev$. 
Particularly, $\{y_{NI}^2=1.2,~\widetilde{m}^2=(20 \tev)^2\}$ leads to a quite exact focusing, 
and so $m_{h_u}^2(t_T)$ is almost invariant under variation of $m_0^2$. 
Again, $m_0^2\approx (8 \tev)^2$ at the GUT scale happens to yield $5 \tev$ stop masses 
and eventually the $126 \gev$ Higgs boson mass. 
Around $m_0^2=(8 \tev)^2$, 
the fine-tuning measure is estimated as 
\dis{
\Delta_{m_0^2}
=\left|\frac{\partial ~{\rm log} ~m_Z^2}{\partial ~{\rm log} ~m_0^2}\right|
~\approx~ 66 ~~~{\rm and}~~~ 306
}
for $\{y_{NI}^2=1.0,~\widetilde{m}^2=(15 \tev)^2\}$ 
and $\{y_{NI}^2=1.2,~\widetilde{m}^2=(20 \tev)^2\}$, respectively. 
They are remarkably small compared to \eq{FTmeasure0}. 
Even for $\{y_{NI}^2=1.0,~\widetilde{m}^2=(10 \tev)^2,~(20 \tev)^2\}$,  
$\Delta_{m_0^2}$ turns out to be just around $65-67$.  
However, it is rather sensitive to $y_{NI}^2$: e.g. for $\{y_{NI}^2=0.8, ~1.2,~\widetilde{m}^2=(15 \tev)^2\}$, 
$\Delta_{m_0^2}$ turns out to be $438$ and $290$, respectively. 
With the hierarchy $\widetilde{m}/m_{1/2}=15-20$, $\Delta_{m_0^2}$ can thus reduce to ${\cal O}(10^2)$ or smaller at one-loop level.\footnote{
Using the public codes, ``SARAH4.2.2'' \cite{SARAH} and ``SPheno3.3.2'' \cite{SPheno} after properly modifying them, one could estimate also other fine-tuning measures at two-loop level: 
e.g. $\Delta_{\alpha}=\{106,32,75,543,71\}$ for $\alpha=\{m_0^2,\widetilde{m}^2,m_{1/2},A_0,\mu\}$, 
when  $y_{NI}=0.8$ and $\widetilde{m}^2= (15 \tev)^2$ with $\alpha_{\rm GUT}\approx 1/25$, $m_{1/2}=(1 \tev)^2$, and $m_0^2=A_0^2=(7 \tev)^2$. 
$A_0$ of $7 \tev$ leads to a relatively large $\Delta_{A_0}$. 
In this case, the stop mixing effect on the Higgs mass is still negligible [$(A_t/\widetilde{m}_t)^2\approx 0.07$] at low energies, yielding $m_H^2\approx(126 \gev)^2$. 
The mass spectra for the neutralino, charginos, and gluino are $\{454 \gev, 505 \gev, 519 \gev, 945 \gev\}$, $\{496 \gev, 944 \gev\}$, and $2.8 \tev$, respectively, 
with $\mu\approx 510 \gev$. 
}

%
As mentioned before, the case that the FP emerges at a scale slightly lower than $t_T$ 
yields a smaller fine-tuning.

%
%
%
\begin{table}[!h]
\begin{center}
\begin{tabular}
{c|ccc||c|ccc}
 $$ & {\footnotesize $\tb=50$}  & {\footnotesize $y_{NI}^2=1.0$}  & {\footnotesize $\widetilde{m}=15 \tev$}  & $$  & {\footnotesize $\tb=50$}  & {\footnotesize $y_{NI}^2=1.2$}  & {\footnotesize $\widetilde{m}=20 \tev$}
\\ \hline
 {\footnotesize ${\bf m_0^2}$} & {\footnotesize $({\bf 9} \tev)^2$}  & {\footnotesize $({\bf 7} \tev)^2$}  & {\footnotesize $({\bf 5} \tev)^2$}  &
{\footnotesize ${\bf m_0^2}$} & {\footnotesize $({\bf 9} \tev)^2$}  & {\footnotesize $({\bf 7} \tev)^2$}  & {\footnotesize $({\bf 5} \tev)^2$} 
\\ \hline
 {\footnotesize $m_{q_3}^2(t_T)$} &  {\footnotesize $(6.3 \tev)^2$}  & {\footnotesize $(4.8 \tev)^2$}  & {\footnotesize $(3.1 \tev)^2$} &
 {\footnotesize $m_{q_3}^2(t_T)$}  & {\footnotesize $(5.9 \tev)^2$}  & {\footnotesize $(4.2 \tev)^2$}  & {\footnotesize $(2.1 \tev)^2$}  
\\
 {\footnotesize $m_{u^c_3}^2(t_T)$} &  {\footnotesize $(5.9 \tev)^2$}  & {\footnotesize $(4.4 \tev)^2$}  & {\footnotesize $(2.9 \tev)^2$}  & {\footnotesize $m_{u^c_3}^2(t_T)$}  & {\footnotesize $(5.5 \tev)^2$}  & {\footnotesize $(3.9 \tev)^2$}  & {\footnotesize $(2.1 \tev)^2$} 
\\
 {\footnotesize ${\bf m_{h_u}^2(t_T)}$} & {\footnotesize $({\bf 1.2} \tev)^2$}  & {\footnotesize $({\bf 0.8} \tev)^2$}  & {\footnotesize $({\bf 0.4} \tev)^2$}  & {\footnotesize ${\bf m_{h_u}^2(t_T)}$}  &  {\footnotesize $({\bf 0.7} \tev)^2$}  & {\footnotesize $({\bf 0.7} \tev)^2$}  & {\footnotesize $({\bf 0.7} \tev)^2$} 
\end{tabular}
\end{center}\caption{Soft squared masses of the stops and Higgs boson at $t=t_T\approx 0.92$ ($Q=5 \tev$) for 
$m_0^2=(9 \tev)^2$, $(7 \tev)^2$, and $(5 \tev)^2$,  
when $\tb=50$ and $m_{1/2}= A_0=1 \tev$ with $\alpha_{G}=1/24$. 
The left [right] four columns correspond to the results of $\{y_{NI}^2=1.0,~\widetilde{m}^2=(15 \tev)^2\}$ [$\{y_{NI}^2=1.2,~\widetilde{m}^2=(20 \tev)^2\}$]. 
}\label{tab:FPn_50}
\end{table}
%
%

Once $\{m_{h_u}^2,m_{h_d}^2\}$ are determined at low energy, $\mu$ should be properly adjusted to give $m_Z^2\approx (91 \gev)^2$ as seen in \eq{extremeCondi}. 
Actually, the RG equation for $\mu$ is decoupled from those of $\{m_{q_3}^2,m_{u_3^c}^2,m_{h_u}^2, {\rm etc.}\}$ at one-loop level, and so 
its evolution does not affect our previous discussions.  
For the case of a small enough $\Delta_{m_0^2}$, 
$\Delta_{\mu}$ ($=|2\frac{\mu^2}{m_Z^2}\frac{\partial m_Z^2}{\partial \mu^2}|$) could become dominant over it \cite{Deltamu,Kowalska}. 
For $m_{h_u}^2(t_T)< (1 \tev)^2$, however, $|\mu|^2$ should be smaller than $(1 \tev)^2$.   
Thus, $\mu^2/m_Z^2$ [$\approx -m_{h_u}^2(t_W)/m_Z^2$] in $\Delta_{\mu}$ is not excessively large ($<100$). 
Moreover, $\Delta_{\mu}$ is closely associated with the mechanism that $\mu$ is generated. 
If $\mu$ is generated at an intermediate scale (rather than the GUT scale), 
$\partial m_Z^2/\partial \mu^2$ can reduce a bit, which further decreases $\Delta_{\mu}$. 



\section{U(1)$^\prime$ mediation and Phenomenological constraints} \label{sec:model}


As seen above, the hierarchy of $\widetilde{m}/m_{1/2}\sim {\cal O}(10)$ is essential 
for a successful FP scenario. 
It can be realized e.g. by employing also the U(1)$^\prime$ mediated SUSY breaking \cite{Zprime}.  
Let us consider the following interaction among vectorlike superfields: 
\begin{eqnarray}
W = (M+\theta^2 F)X X^c + y_1 X\Phi\Psi^c + y_2  X^c\Phi^{c} \Psi 
+ M_{\Phi}\Phi \Phi^{c} + M_\psi \Psi\Psi^c ,
\end{eqnarray}
where $M$ and $F$ denotes the scalar and $F$-components of a spurion superfield ($\Sigma$) parametrizing SUSY breaking effect.
$M_{\Phi,\Psi}$ ($\sim M_{G}$) and $y_{1,2}$ are dimensionful and dimensionless parameters, respectively. 
For the above superpotential, one can assign e.g. U(1)$_R$ charges of 2 and 1 to $\Sigma$ and $\{\Phi,\Phi^c;\Psi,\Psi^c\}$, respectively. 
$\{X,X^c\}$, which are neutral under U(1)$_R$, 
play the role of the messenger for SUSY breaking effects on the MSSM sector. 
While $\{X, X^c\}$ are U(1)$^\prime$ charged but SM singlet superfields,  
$\{\Phi, \Phi^{c}\}$ are superfields carrying both U(1)$^\prime$ and SM gauge charges. 
$\{\Psi,\Psi^c\}$ carry only SM gauge quantum numbers. 
In the U(1)$^\prime$ mediated SUSY breaking scenario \cite{Zprime},  
the U(1)$^\prime$ gaugino mass ($\equiv M_{\widetilde Z^\prime}$) is of order $(g_{\widetilde Z^\prime}^2/16\pi^2)F/M$. 
On the other hand, the soft squared masses of U(1)$^\prime$ charged scalars, i.e., the first and second generations of sfermions in our case are given by $M_{\widetilde Z^\prime}$, 
$\widetilde m^2 \sim (q_i^2g_{\widetilde Z^\prime}^2/16\pi^2) M_{\widetilde Z^\prime}^2$. 
$m_0^2$ can be induced just through the ordinary gravity mediated SUSY breaking effect, 
which is always there. 
Thus, the soft squared masses for the third generation of sfermions are given by $m_0^2$. 

Since the SM charged superfields have Yukawa interactions with the messengers, 
the threshold correction to the wave function renormalization  for $\Psi^c$ has the following form:  
\begin{eqnarray}
\Delta Z_{\Psi^c} \sim \frac{y_1^2}{16\pi^2} {\rm log} |M+\theta^2 F|^2 . 
\end{eqnarray}
It contributes to the MSSM gaugino masses: 
\bea
\frac{m_{1/2}}{g_G^2} \sim  -\frac{1}{8\pi^2} {\rm Tr}\left[T_G^2(\Psi^c)\right] \left[{\rm log} Z_{\Psi^c}\right]\bigg|_F
={\cal O}\left(\frac{y_1^2}{(16\pi^2)^2} \frac{F}{M}\right)
={\cal O}\left(\frac{M_{\widetilde Z^\prime}}{16\pi^2}\right).
\eea
We regard it as the dominant contribution to the MSSM gaugino masses. 
Hence, in this setup, we can achieve the desired hierarchy, $\widetilde m/m_{1/2}\sim {\cal O}(4\pi)$.

According to the ``effective SUSY'' (or ``more minimal SUSY''), the masses of the first two generations of sfermions are required to be about $5$--$20 \tev$ in order to avoid the SUSY flavor and SUSY $CP$ problems, while the third ones and gauginos are lighter than $1 \tev$ for naturalness of the Higgs boson \cite{effSUSY}. 
In our case, the third generations of sfermions are heavier than $1 \tev$, but the naturalness problem can be addressed depending on the FP scenario. 
As in the effective SUSY, the hierarchically heavy masses for the first two generations of sfermions ($15$--$20 \tev$) with $CP$ violating phases of ${\cal O}(0.1)$ 
can solve the SUSY flavor and SUSY $CP$ problems.  
In Ref.~\cite{Natural/2-loop}, it was pointed out that such heavy masses for the first two generations of sfermions drive the stop mass squared too small or even negative at the EW scale via RG evolutions. 
As seen in Tables~\ref{tab:FPn_5} and \ref{tab:FPn_50}, however, such a thing does not occur. 
It is because the gluino mass is quite heavy in our case. 
Moreover, the initial value of stop squared masses at the GUT scale, $m_0^2$ can be quite large 
without a serious fine-tuning only if $m_{h_u}^2(t)$ is well focused near the stop mass scale.    


Since all the sfermions are very heavy in this model, the pair annihilation cross section of the lightest neutralino is quite suppressed, and so it would overclose the Universe. 
However, this problem could be resolved, e.g. if a sufficient amount of entropy is somehow produced after thermal freeze-out of the neutralino \cite{RHnuFP}. 
In this paper, we do not discuss this issue in detail.   
Instead, let us discuss phenomenological constraints coming from flavor violations in more detail. 

In the squark mass matrix, the diagonal components, $(1,1)$ and $(2,2)$, are almost degenerate with a squared mass of $(15$--$20 \tev)^2$, e.g., by the U(1)$^\prime$ SUSY breaking mediation, while 
the $(3,3)$ is filled dominantly by the gravity mediation effect, which is quite suppressed compared to the $(1,1)$ and $(2,2)$ components. 
In the other components, nonzero values can be generated by a U(1)$^\prime$ breaking effect. (We do not specify a U(1)$^\prime$ breaking mechanism here.) 
After diagonalization in the fermionic quarks sector, $(1,2)$, $(2,1)$, and $(i,3)$, $(3,i)$ can be induced after U(1)$^\prime$ breaking. 

The $(1,2)$ and $(2,1)$ components affect, e.g., $K$-$\bar K$ mixing.   
The amplitude of $K$-$\bar K$ mixing by the squark mixing is roughly estimated as \cite{book2,FV}
\dis{ \label{KKbar}
{\cal M}_{K\bar{K}}\approx
\frac{4\alpha_3^2}{\widetilde m_q^2}
\left(\frac{\Delta\widetilde m_q^2}{\widetilde m_q^2}\right)^2 ,
}
where $\widetilde m_q^2\approx (20 \tev)^2$, and  $\Delta\widetilde m_q^2$ denotes the off-diagonal component of the squark mass matrix. 
Note that RG runnings of the heavy masses for the first two generations of sfermions are negligible \cite{Natural/2-loop,Natural/splitZprime}, and so 
their low energy values are almost the same as those at the GUT scale.  
Since the SM still explains the observed data well, 
\eq{KKbar} should be smaller than the SM prediction,
${\cal M}_{K\bar{K}}^{\rm SM}\approx\alpha_2^2\sin^2\theta_c\cos^2\theta_c
(m_c^2/M_W^4)$, where $\theta_c$ stands for the Cabibbo mixing angle. 
The condition ${\cal M}_{K\bar{K}}\ll{\cal M}_{K\bar{K}}^{\rm SM}$ yields
\dis{
\left(\frac{\Delta \widetilde m_q^2}{\widetilde m_q^2}\right)\ll 1.6\times 10^{-1}\times 
\left(\frac{\widetilde m_q}{20 \tev}\right) .
}
If the mixing among the d-type quarks is given fully by the CKM (or a similar order mixing matrix) and the elements induced by gravity mediation are of order TeV$^2$, this constraint can be satisfied.\footnote{In fact, even $\widetilde{m}_q^2\approx (10 \tev)^2$ is enough to avoid the SUSY flavor and SUSY $CP$ problems in the quark sector \cite{Delgado}.} 

Unlike the quark sector, the lepton sector requires large mixing to explain the observed neutrino oscillations.
Thus, although $(1,1)$ and $(2,2)$ components of the slepton mass matrices acquire very large squared masses [$\approx (15$--$20 \tev)^2$] from the U(1)$^\prime$ mediation effect,    
other components can also receive large squared masses after diagonalization of the fermion mass matrices.
Nonzero off-diagonal components in the slepton matrix can induce 
lepton flavor violations (LFV), which is absent in the SM. 
The branching ratio for $\mu^-\to e^-\gamma$ by such a slepton mixing is estimated as \cite{FV}  
\dis{
&\frac{{\rm BR}(\mu^-\to e^-\gamma)}{{\rm BR}(\mu^-\to e^-\nu_\mu\bar{\nu}_e)}
=\frac{12\pi\alpha^3}{G_F^2\widetilde{m}^4_l}
\left\{\left|I_3(x)\left(\delta^l_{21}\right)_{LL}
+\frac{M_{\tilde\gamma}}{m_\mu}I_1(x)\left(\delta^l_{21}\right)_{LR}\right|^2
+L\leftrightarrow R\right\}
\\
&~~ \approx 6.7\times 10^{-13}\times\left[\frac{(20 \tev)^4}{\widetilde{m}^4_l}\right]
\left\{\left|\frac{1}{12}\left(\delta^l_{21}\right)_{LL}
+\frac{M_{\tilde\gamma}}{2m_\mu}\left(\delta^l_{21}\right)_{LR}\right|^2
+L\leftrightarrow R\right\} , 
}
where the functions of $x$, $I_{3}(x)$ and $I_1(x)$ approach to $1/12$ and $1/2$, respectively, for $x\equiv M_{\tilde\gamma}^2/\widetilde{m}_e^2\ll 1$. 
$\widetilde{m}_l$ is the mass of the first or second generation of SU(2)$_L$ doublet (i.e., LH) slepton.  
$\left(\delta^l_{21}\right)_{LR}$ is associated with the $A$-term vertex proportional to a very small Yukawa coupling. 
It is at most of order $m_\mu/\widetilde{m}_e$, 
which suppresses the second term, because  
the photino mass $M_{\tilde\gamma}$ would be smaller than $1 \tev$ in our case. 
This process is possible through, e.g., the $\tilde{\nu}_{1,2}$-chargino and $\tilde{e}_{1,2}$-neutralino loops.
Even if the slepton mixing $\left(\delta^l_{21}\right)_{LL}$ is of order unity,  
sleptons of 20 TeV are heavy enough to meet the current bound, ${\rm BR}(\mu^-\to e^-\gamma) < 5.7\times 10^{-13}$ \cite{muegamma}.

Similarly, such heavy slepton masses suppress also $\tau^-\to e^-\gamma$ [${\rm BR}(\tau^-\to e^-\gamma) < 3.3\times 10^{-8}$ \cite{PDG}]
and $\tau^-\to \mu^-\gamma$ [${\rm BR}(\tau^-\to \mu^-\gamma) < 4.4\times 10^{-8}$], 
which are actually much less stringent,  
because they are still involved in those processes.
Even if the first two generations of sleptons are quite heavy, however, 
%
$\tau$ can still decay with a sizable rate through the $\tilde{\nu}_{3L}$-chargino and $\tilde{e}_{3L}$-neutralino loops without a slepton mixing insertion, 
provided that the $\tau$--$e$ or $\tau$--$\mu$ mixing in the fermion sector is large \cite{Hisano}.
So it is desirable to assume that the PMNS matrix 
comes dominantly from the neutrino sector \cite{Natural/splitZprime}, 
only if the first two generations of sleptons are quite heavy.  
Then additional large off-diagonal components of sneutrino mass matrix, 
which are induced after diagonalization of the neutrino mass matrix, can suppress the unwanted $\tau^-\to e^-\gamma$ and $\tau^-\to \mu^-\gamma$.  
%

Now we propose a model, in which the PMNS matrix results from mixing of the neutrino sector. 
Let us introduce extra singlet fields.  
Their charge assignments under U(1)$^\prime$ and U(1)$_R$ are listed in Table~\ref{tab:GC}.
%
%
%
\begin{table}[!h]
\begin{center}
\begin{tabular}
{c|ccc|ccc}
Superfields ~&~ $l_{1,2}$  ~&~ $e^c_{1,2}$  ~&~ $l_3,e^c_3,\nu^c_{1,2,3}$  
 ~&~ $S_{1,2}$ ~&~ $S_{1,2}^c$ ~&~ $Z_{1,2}$    
\\ \hline
U(1)$^\prime$ & $\pm 2$ & $\mp 2$ & $0$ & $\mp 2$  & $\pm 1$  & $\pm 1$ 
\\
U(1)$_{R}$ & $1$ & $1$ & $1$ & $1$ & $1$ & $0$ 
\end{tabular}
\end{center}\caption{
U(1)$^\prime$ and U(1)$_R$ charges for various superfields. The MSSM Higgs doublets are neutral under both symmetries. The subscripts of the MSSM superfields are family indices. 
U(1)$^\prime$ is assumed to be broken by nonzero VEVs of $\widetilde{Z}_{1,2}$ around the GUT scale.
}\label{tab:GC} 
\end{table}
%
%
%
One can see that the charged lepton mass matrix should have a diagonal form at the renormalizable level because of the U(1)$^\prime$ and U(1)$_R$ symmetries.  
Through the U(1)$^{\prime}$ mediated SUSY breaking mechanism, 
sfermions with nonzero U(1)$^\prime$ charges receive quite heavy soft masses. 
Hence, as discussed above, LFV can adequately be suppressed by U(1)$^\prime$.
Note that the RH neutrinos, $\nu^c_{1,2,3}$ carry only the U(1)$_R$ [and U(1)$_{B-L}$] charge[s]. 
So they can freely be mixed. 
Note that the mixing in the RH (s)neutrino sector is almost irrelevant to LFV, 
while RH neutrinos' mixing still contributes to the PMNS matrix.

The superpotential of the neutrino sector consistent with  U(1)$^\prime\times$U(1)$_R$ 
is written as 
\dis{ \label{W_N}
W_N=& \sum_{i=1,2,3}\left[y_{\nu}^il_3h_u\nu^c_i +\frac12M^{ij}\nu_i^c\nu_j^c 
+ \left(\lambda_1^{i} Z_2 S_1^c + \lambda_2^{i} Z_1 S_2^c\right)\nu^c_i\right] 
\\
&+ \sum_{k=1,2}\left[y_S^k l_{k}h_uS_k + \lambda_Z^k Z_kS_kS_k^c\right] 
+ M_SS_1S_2 + M_{S^c}S_1^cS_2^c ,
}
where $M^{ij}$ ($\{M_S,M_{S^c}\}$) denotes dimensionful parameters of order $10^{14} \gev$ or smaller ($10^{16} \gev$ or smaller), 
while $y$s and $\lambda$s are dimensionless ones. 
[$M^{ij}$ breaks U(1)$_{B-L}$.] 
In terms of \eq{W_N}, $N^c$ in \eq{RHnu} can be identified as 
$(y_{\nu}^1\nu^c_1+y_{\nu}^2\nu^c_2+y_{\nu}^3\nu^c_3)/\sqrt{\sum_{i}(y_{\nu}^i)^2}$, 
and $y_N$ as $\sqrt{(y_{\nu}^1)^2+(y_{\nu}^2)^2+(y_{\nu}^3)^2}$. 
The other two components orthogonal to $N^c$ have no direct couplings to the MSSM lepton doublets. 
They obtain such couplings via the mediation of $\{S_{1,2},S_{1,2}^c\}$ 
after $\widetilde{Z}_{1,2}$ get GUT scale VEVs, breaking U(1)$^\prime$, and $\{S_{1,2},S_{1,2}^c\}$ are integrated out. 
We assume that the resulting effective Dirac Yukawa couplings are somewhat smaller than $y_N$. 
The sizable (effective) Dirac Yukawa couplings could radiatively generate 
the mixing soft mass squareds such as $(\Delta\widetilde{m}_{31})_{LL}$, $(\Delta\widetilde{m}_{32})_{LL}$, etc. for sneutrinos 
via the RH neutrino-Higgsino loops above the seesaw scale.\footnote{If the U(1)$^\prime$ breaking scale and the mass scale of $S_{1,2}^{(c)}$ are lower than the seesaw scale, 
they are not radiatively generated at all even with sizable Dirac neutrino Yukawa couplings.} 
As discussed above, however, such mixing terms cannot give rise to sizable LFV,  
because the heavy soft masses for sleptons should always be involved there.      
%
After integrating out the RH neutrinos $\nu^c_{1,2,3}$, the general results of the type-I seesaw mechanism can eventually be reproduced. 
Unlike the charged lepton sector,  the neutrinos can thus fully be mixed below the seesaw scale, yielding the desired form of the PMNS matrix in principle. 
In a similar way, one can achieve the CKM mixing of the quarks by introducing extra vectorlike quarks at the GUT scale, which play the role of the mediators  $\{S_{1,2},S_{1,2}^c\}$. 
However, the absence of the extra vectorlike charged leptons guarantees the almost diagonal mass matrix for the SM charged leptons even at low energies.

\section{Conclusion}  \label{sec:conclusion}


According to the recent analysis based on three-loop calculations, 
the radiative correction by $5 \tev$ stop masses can support the $126 \gev$ Higgs 
mass without a large stop mixing effect. 
The $5 \tev$ stop decoupling scale is much higher than the FP scale 
determined in the original FP scenario. 
As a result, $m_{h_u}^2$ evaluated at low energy becomes sensitive to $m_0^2$ 
chosen at the GUT scale, and so to the low energy value of stop mass, unlike the original FP scenario. 
Moreover, the present high gluino mass bound ($\gtrsim 1.4 \tev$) results in a too large negative $m_{h_u}^2$ at low energy, 
which gives rise to a serious fine-tuning problem in the MSSM Higgs sector. 
   
In this paper, we have discussed how the location of the FP changes 
under various variations of parameters. 
In particular, we noted that the FP can move to the desirable location 
under increases of {\it both} the Yukawa coupling of a superheavy RH neutrino to the Higgs, 
and the masses of the first and second generations of sfermions. 
On the other hand, the ``$\lambda$ coupling'' in the NMSSM should be more suppressed than $0.1$ to be consistent with the FP scenario, if it is introduced.     

We have shown that an order one Dirac Yukawa coupling ($\sim 1.0$) of the superheavy RH neutrino ($\sim 10^{14} \gev$) at the seesaw scale 
can move the FP to the desired stop decoupling scale, 
and two-loop gauge interactions by the hierarchically heavy masses ($15-20 \tev$) 
of the first two generations of sfermions can effectively compensate the heavy gluino effects in the RG evolution of $m_{h_u}^2$. 
Here, we set the U(1)$_R$ breaking soft parameters,  $m_{1/2}=A_0=1 \tev$, at the GUT scale. 
The gaugino mass unification is maintained in this setup.
Such heavy masses of the RH neutrino and the first two generations of sfermions 
can also provide a natural explanation of the small active neutrino mass via the seesaw mechanism, and suppress the flavor violating processes in SUSY models. 
At the new location of the FP, $m_{h_u}^2$ can be insensitive to $m_0^2$ or trial heavy stop squared masses, remarkably improving the naturalness of the small EW scale.
Under this setup, the $126 \gev$ Higgs mass can be naturally explained by an accidentally selected 
$m_0^2$ of about $(8 \tev)^2$, which gives $5 \tev$ stop mass at low energy. 
%

\acknowledgments

B.K. thanks Department of Physics and Astronomy in Rutgers University 
for the hospitality during his visit to Rutgers University. 
B.K. is supported by 
the National Research Foundation of Korea (NRF) funded by the Ministry of Education, Grant No. 2013R1A1A2006904, and also in part 
by Korea Institute for Advanced Study (KIAS) grant funded by the Korean government.
C.S.S. is supported in part by DOE Awards No. DOE-SC0010008, No. DOE-ARRA-SC0003883, and No. DOE-DE-SC0007897.


\section{Appendix} \label{sec:Appendix}



In the Appendix, we present the full RG equations utilized in our analyses and 
some semianalytic solutions on which the discussions in the main text are based. 
The notations here follow those of the main text of this paper.

\subsection{The full RG equations}

The RG equations for the gauge couplings, $g_{3,2,1}$ 
and gaugino masses, $M_{3,2,1}$ are integrable.  
The RG solutions for them are given by \cite{book}  
\begin{eqnarray} \label{gaugeSol}
g_i^2(t)=\frac{g_0^2}{1-\frac{g_0^2}{8\pi^2}b_i(t-t_0)}~,
~~~~ {\rm and}~~~~~ \frac{M_i(t)}{g_i^2(t)}=\frac{m_{1/2}}{g_0^2} ~, 
\end{eqnarray}
where $b_i$ ($i=3,2,2$) denotes the beta function coefficients for the case of the MSSM field contents, $(b_3,b_2,b_1)=(-3,1,\frac{33}{5})$. 
$t$ parametrizes the renormalization scale $Q$, $t-t_0={\rm log}\frac{Q}{M_{G}}$.   
The relevant superpotential in this paper is  
\dis{ \label{apdxSuperPot}
W\supset y_tq_3h_uu^c_3 + y_bq_3h_dd^c_3 + y_\tau l_3h_de_3^c 
+ y_N l_3h_uN^c + \frac{1}{2}M_NN^cN^c + \mu h_uh_d,   
}
where $q_3$ ($l_3$) and $\{u^c_3,d^c_3\}$ ($e^c_3$) stand for the third generations of quark (lepton) doublet and singlets.  
The Majoran mass of the RH neutrino $N^c$ is assumed to be $M_N\approx 2\times 10^{14} \gev$. 
Thus, below the energy scale of $M_N$, the RH neutrino $N^c$ is decoupled from dynamics. 
The one-loop RG equations for the above renormalizable couplings are given by   
\bea
&&8\pi^2\frac{dy_t^2}{dt}=y_t^2\left[6y_t^2+y_b^2+y_N^2-\frac{16}{3}g_3^2-3g_2^2-\frac{13}{15}g_1^2\right] ,
\\
&&8\pi^2\frac{dy_b^2}{dt}=y_b^2\left[y_t^2+6y_b^2+y_\tau^2-\frac{16}{3}g_3^2-3g_2^2-\frac{7}{15}g_1^2\right] ,
\\
&&8\pi^2\frac{dy_\tau^2}{dt}=y_\tau^2\left[3y_b^2+4y_\tau^2+y_N^2-3g_2^2-\frac{9}{5}g_1^2\right] ,
\\
&&8\pi^2\frac{dy_N^2}{dt}=y_N^2\left[3y_t^2+y_\tau^2+4y_N^2-3g_2^2-\frac{3}{5}g_1^2\right] ,  
\\
&&8\pi^2\frac{d\mu^2}{dt}=\mu^2\left[3y_t^2+3y_b^2+y_\tau^2+y_N^2-3g_2^2-\frac{3}{5}g_1^2\right] ,
\eea
and the RG equations of the $A$-term coefficients corresponding to the Yukawa couplings of \eq{apdxSuperPot} are  
\bea
8\pi^2\frac{dA_t}{dt}&=&6y_t^2A_t+y_b^2A_b+y_N^2A_N-\frac{16}{3}g_3^2M_3-3g_2^2M_2-\frac{13}{15} g_1^2M_1 ,
\\
8\pi^2\frac{dA_b}{dt}&=&y_t^2A_t+6y_b^2A_b+y_\tau^2A_\tau-\frac{16}{3}g_3^2M_3-3g_2^2M_2-\frac{7}{15} g_1^2M_1 ,
\\
8\pi^2\frac{dA_\tau}{dt}&=&3y_b^2A_b+4y_\tau^2A_\tau+y_N^2A_N-3g_2^2M_2-\frac{9}{5} g_1^2M_1 ,
\\
8\pi^2\frac{dA_N}{dt}&=&3y_t^2A_t+y_\tau^2A_\tau+4y_N^2A_N-3g_2^2M_2-\frac{3}{5} g_1^2M_1 .
\eea
Below the scale of $M_N$, the RG evolutions of $y_N$ and $A_N$ become frozen, 
and they should be decoupled from the above equations.  

The RG evolutions for the soft squared masses are governed by the following equations: 
\bea
16\pi^2\frac{dm_{h_u}^2}{dt}&=&6y_t^2\left(X_t+A_t^2\right)+2y_N^2\left(X_N+A_N^2\right)-6g_2^2M_2^2-\frac65 g_1^2M_1^2
+\frac{\widetilde{m}^2}{4\pi^2}\left[6g_2^4+\frac65g_1^4\right] , 
\label{} \qquad~ \\
16\pi^2\frac{dm_{u^c_3}^2}{dt}&=&4y_t^2\left(X_t+A_t^2\right)-\frac{32}{3}g_3^2M_3^2-\frac{32}{15} g_1^2M_1^2 
+\frac{\widetilde{m}^2}{4\pi^2}\left[\frac{32}{3}g_3^4+\frac{32}{15} g_1^4 \right] ,
\label{} \\
16\pi^2\frac{dm_{q_3}^2}{dt}&=&2y_t^2\left(X_t+A_t^2\right) + 2y_b^2\left(X_b+A_b^2\right)  -\frac{32}{3}g_3^2M_3^2-6g_2^2M_2^2-\frac{2}{15} g_1^2M_1^2 
\label{} \\
&& ~~ +\frac{\widetilde{m}^2}{4\pi^2}\left[\frac{32}{3}g_3^4+6g_2^4+\frac{2}{15} g_1^4\right] ,
\nonumber  \\
16\pi^2\frac{dm_{h_d}^2}{dt}&=&6y_b^2\left(X_b+A_b^2\right)+2y_\tau^2\left(X_\tau+A_\tau^2\right)
-6g_2^2M_2^2-\frac65 g_1^2M_1^2
+\frac{\widetilde{m}^2}{4\pi^2}\left[6g_2^4+\frac65g_1^4\right] , 
\label{} \\
16\pi^2\frac{dm_{d^c_3}^2}{dt}&=&4y_b^2\left(X_b+A_b^2\right)-\frac{32}{3}g_3^2M_3^2-\frac{8}{15} g_1^2M_1^2 
+\frac{\widetilde{m}^2}{4\pi^2}\left[\frac{32}{3}g_3^4+\frac{8}{15} g_1^4 \right] ,
\label{} \\
16\pi^2\frac{dm_{e^c_3}^2}{dt}&=&4y_\tau^2\left(X_\tau+A_\tau^2\right)  -\frac{24}{5} g_1^2M_1^2 
+\frac{\widetilde{m}^2}{4\pi^2}\left[\frac{24}{5} g_1^4\right] ,
  \\
16\pi^2\frac{dm_{l_3}^2}{dt}&=&2y_\tau^2\left(X_\tau+A_\tau^2\right)+2y_N^2\left(X_N+A_N^2\right) -6g_2^2M_2^2-\frac{6}{5} g_1^2M_1^2 
+\frac{\widetilde{m}^2}{4\pi^2}\left[6g_2^4+\frac{6}{5} g_1^4\right] ,  
 \\
16\pi^2\frac{dm_{N^c}^2}{dt}&=&4y_N^2\left(X_N+A_N^2\right) ,
\label{} 
\eea
where $X_t$, $X_b$, $X_\tau$, and $X_N$ are defined as $X_t\equiv m_{h_u}^2+m_{u^c_3}^2+m_{q_3}^2$, $X_b\equiv m_{h_d}^2+m_{d^c_3}^2+m_{q_3}^2$, $X_\tau\equiv m_{h_d}^2+m_{e^c_3}^2+m_{l_3}^2$, and $X_N\equiv m_{h_u}^2+m_{N^c}^2+m_{l_3}^2$, respectively. 
The $\widetilde{m}^2$ terms denote the contributions coming from the two-loop gauge interactions by the first and second generations of sfermions, which are assumed to be hierarchically heavier than the third ones. 
The RG running of $\widetilde{m}^2$ is negligible \cite{Natural/2-loop,Natural/splitZprime}, and so 
its low energy value is almost the same as that at the GUT scale.
Here we suppose a universal soft mass for the first two generations of sfermions, which eliminates the contributions by the ``$D$-term'' potential from the above equations.  
Since these effects are comparable to the one-loop gaugino mass terms, 
we take them into account.   
$m_N^2$ and $X_N$ as well as $y_N$ and $A_N$ are dropped out from the above equations 
below $Q= M_N$.

\subsection{Semianalytic RG solutions}

Let us present our semianalytic solutions to the RG equations. 
When ${\rm tan}\beta$ is small enough and the RH neutrino is decoupled, 
the RG evolutions of the soft mass parameters, 
$m_{h_u}^2$, $m_{u^c_3}^2$, $m_{q_3}^2$, and $A_t$ are approximately simplified as 
\begin{eqnarray}
16\pi^2\frac{dm_{h_u}^2}{dt}&=&6y_t^2\left(X_t+A_t^2\right)-6g_2^2M_2^2-\frac65 g_1^2M_1^2
+\frac{\widetilde{m}^2}{4\pi^2}\left[6g_2^4+\frac65g_1^4\right] , 
\label{apdxRG1} \\
16\pi^2\frac{dm_{u^c_3}^2}{dt}&=&4y_t^2\left(X_t+A_t^2\right)-\frac{32}{3}g_3^2M_3^2-\frac{32}{15} g_1^2M_1^2 
+\frac{\widetilde{m}^2}{4\pi^2}\left[\frac{32}{3}g_3^4+\frac{32}{15} g_1^4 \right] ,
\label{apdxRG2} \\
16\pi^2\frac{dm_{q_3}^2}{dt}&=&2y_t^2\left(X_t+A_t^2\right)-\frac{32}{3}g_3^2M_3^2-6g_2^2M_2^2-\frac{2}{15} g_1^2M_1^2 
+\frac{\widetilde{m}^2}{4\pi^2}\left[\frac{32}{3}g_3^4+6g_2^4+\frac{2}{15} g_1^4\right] ,
\qquad~~ \label{apdxRG3} \\
8\pi^2\frac{dA_t}{dt}&=&6y_t^2A_t-\frac{16}{3}g_3^2M_3-3g_2^2M_2-\frac{13}{15} g_1^2M_1 
\equiv 6y_t^2A_t - G_A . 
\label{apdxRG4}
\end{eqnarray} 
Summation of Eqs.~(\ref{apdxRG1}), (\ref{apdxRG2}), and (\ref{apdxRG3}) yields the RG equation for $X_t$:
\dis{ \label{apdxX}
\frac{dX_t}{dt} = \frac{3y_t^2}{4\pi^2}\left(X_t + A_t^2\right)
-\frac{1}{4\pi^2} G_X^2 .
%
}
In Eqs.~(\ref{apdxRG4}) and (\ref{apdxX}), $G_A$ and $G_X^2$ are defined as 
\begin{eqnarray}
&&\qquad\qquad\quad~~ G_A(t)\equiv 
\left(\frac{m_{1/2}}{g_0^2}\right)\left[\frac{16}{3}g_3^4+3g_2^4+\frac{13}{15}g_1^4\right] ,
\\
&&G_X^2(t)\equiv 
\left(\frac{m_{1/2}}{g_0^2}\right)^2\left[\frac{16}{3}g_3^6+3g_2^6+\frac{13}{15}g_1^6\right]-\frac{\widetilde{m}^2}{4\pi^2}\left[\frac{16}{3}g_3^4+3g_2^4+\frac{13}{15}g_1^4\right] ,
\end{eqnarray}
respectively, assuming 
$\frac{M_i(t)}{g_i^2(t)}=\frac{m_{1/2}}{g_0^2}$ ($i=3,2,1$).

The solutions of $A_t$ and $X_t$ are given by 
\begin{eqnarray}
&&\qquad~~ A_t(t)=e^{\frac{3}{4\pi^2}\int^t_{t_0}dt^\prime y_t^2} 
\left[A_0-\frac{1}{8\pi^2}\int^t_{t_0}dt^\prime G_A
e^{\frac{-3}{4\pi^2}\int^{t'}_{t_0}dt^{\prime\prime} y_t^2}
\right] , 
\label{solA}
\\
&&X_t(t)=e^{\frac{3}{4\pi^2}\int^t_{t_0}dt^\prime y_t^2}
\left[X_0+\int^t_{t_0}dt^\prime 
\left(\frac{3}{4\pi^2}y_t^2A_t^2-\frac{1}{4\pi^2}G_X^2\right)
e^{\frac{-3}{4\pi^2}\int^{t'}_{t_0}dt^{\prime\prime}y_t^2}
\right] ,
\label{solX}
\end{eqnarray}
where $A_0$ and $X_0$ denote the GUT scale values of $A_t$ and $X_t$, $A_0\equiv A_t(t=t_0)$, and $X_0\equiv X_t(t=t_0)=m_{h_u0}^2+m_{u^c_30}^2+m_{q_30}^2$.  

With Eqs.~(\ref{apdxX}), (\ref{solA}), and (\ref{solX}), one can solve Eqs. (\ref{apdxRG1}), (\ref{apdxRG2}), and (\ref{apdxRG3}):
\begin{eqnarray}
&&m_{h_u}^2(t)=m_{h_u0}^2+\frac{X_0}{2}\left[e^{\frac{3}{4\pi^2}\int^t_{t_0}dt^\prime y_t^2}-1\right]
+\frac12 F(t)
\nonumber \\
&&\qquad -\left(\frac{m_{1/2}}{g_0^2}\right)^2\left[\frac32\left\{g_2^4(t)-g_0^4\right\}
+\frac{1}{22}\left\{g_1^4(t)-g_0^4\right\}\right]
\label{apdxSol1} \\
&&\qquad +\left(\frac{\widetilde{m}^2}{4\pi^2}\right)\left[3\left\{g_2^2(t)-g_0^2\right\}
+\frac{1}{11}\left\{g_1^4(t)-g_0^4\right\}\right] ,
\nonumber \\
&&m_{u^c_3}^2(t)=m_{u^c_30}^2+\frac{X_0}{3}\left[e^{\frac{3}{4\pi^2}\int^t_{t_0}dt^\prime y_t^2}-1\right]
+\frac13 F(t)
\nonumber \\
&&\qquad +\left(\frac{m_{1/2}}{g_0^2}\right)^2\left[\frac89\left\{g_3^4(t)-g_0^4\right\}
-\frac{8}{99}\left\{g_1^4(t)-g_0^4\right\}\right] 
\label{apdxSol2}  \\
&&\qquad - \left(\frac{\widetilde{m}^2}{4\pi^2}\right)\left[\frac{16}{9}\left\{g_3^2(t)-g_0^2\right\}
-\frac{16}{99}\left\{g_1^2(t)-g_0^2\right\}\right] ,
\nonumber \\ 
&&m_{q_3}^2(t)=m_{q_30}^2+\frac{X_0}{6}\left[e^{\frac{3}{4\pi^2}\int^t_{t_0}dt^\prime y_t^2}-1\right]
+\frac16 F(t)
\nonumber \\
&&\qquad +\left(\frac{m_{1/2}}{g_0^2}\right)^2\left[\frac89\left\{g_3^4(t)-g_0^4\right\}
-\frac32\left\{g_2^4(t)-g_0^4\right\}-\frac{1}{198}\left\{g_1^4(t)-g_0^4\right\}\right] 
\label{apdxSol3} \\
&&\qquad -\left(\frac{\widetilde{m}^2}{4\pi^2}\right)\left[\frac{16}{9}\left\{g_3^2(t)-g_0^2\right\}
-3\left\{g_2^2(t)-g_0^2\right\}-\frac{1}{99}\left\{g_1^2(t)-g_0^2\right\}\right] ,
\nonumber 
\end{eqnarray}
where $F(t)$ is defined as 
\dis{ \label{apdxF}
&\qquad F(t)\equiv e^{\frac{3}{4\pi^2}\int^t_{t_0}dt^\prime y_t^2} \int^t_{t_0}dt^\prime ~\frac{3}{4\pi^2}y_t^2A_t^2 ~e^{\frac{-3}{4\pi^2}\int^{t'}_{t_0}dt^{\prime\prime}y_t^2}
\\
&-\frac{1}{4\pi^2} \left[e^{\frac{3}{4\pi^2}\int^t_{t_0}dt^\prime y_t^2} \int^t_{t_0}dt^\prime ~G_X^2 ~e^{\frac{-3}{4\pi^2}\int^{t'}_{t_0}dt^{\prime\prime}y_t^2}
-\int^t_{t_0}dt^\prime~G_X^2 \right] .
}
Note that $F(t)$ in \eq{apdxF} is independent of the initial values for the squared masses, $m_{h_u0}^2$, $m_{u^c_30}^2$, and $m_{q_30}^2$.
Using Eqs.~(\ref{gaugeSol}), one can obtain the following useful results: 
\begin{eqnarray}
&& \int^t_{t_0}dt^\prime g_i^2M_i^2=\frac{4\pi^2}{b_i}\left(\frac{m_{1/2}}{g_0^2}\right)^2\left\{g_i^4(t)-g_0^4\right\} ,
\\
&& \int^t_{t_0}dt^\prime g_i^2M_i=\frac{8\pi^2}{b_i}\left(\frac{m_{1/2}}{g_0^2}\right)\left\{g_i^2(t)-g_0^2\right\} ,
\\
&& \int^t_{t_0}dt^\prime g_i^4=\frac{8\pi^2}{b_i}\left\{g_i^2(t)-g_0^2\right\} .   
\end{eqnarray} 




\end{document}